\documentclass[aps,prb,twocolumn,groupedaddress,showpacs,superscriptaddress,amssymb,amsmath,noeprint]{revtex4-2}

\usepackage{mathrsfs}
\usepackage{graphicx}
\usepackage{dcolumn}
\usepackage{bm,xcolor,alphabeta,notes2bib}
\usepackage{hyperref}
\usepackage{multirow} 

\usepackage{footmisc}
\usepackage{newunicodechar}
\newunicodechar{−}{-}
\usepackage[normalem]{ulem}

\newcommand{\tr}{\mathrm{tr}}
\newcommand{\bH}{{\bf H}}
\newcommand{\cF}{\mathcal{F}}

\newcommand{\cK}{\mathcal{K}}

\begin{document}

\preprint{APS/123-QED}

\title{Dissipative spectral form factor for elliptic Ginibre unitary ensemble and applications}

\author{Sunidhi Sen}
\email{sensunidhi96@gmail.com}
\affiliation{Department of Physics, Shiv Nadar Institution of Eminence, Gautam Buddha Nagar, Uttar Pradesh 201314, India}
\author{Santosh Kumar}
\email{Deceased.}
\affiliation{Department of Physics, Shiv Nadar Institution of Eminence, Gautam Buddha Nagar, Uttar Pradesh 201314, India}
\author{Ayana Sarkar}
\email{Ayana.Sarkar@usherbrooke.ca}
\affiliation{Institut Quantique $\&$ Departement de Physique,Universit\'{e} de Sherbrooke, J1K 2R1, Quebec, Canada}
\author{Manas Kulkarni}
\email{manas.kulkarni@icts.res.in}
\affiliation{International Centre for Theoretical Sciences, Tata Institute of Fundamental Research, Bangalore 560089, India}
 

\begin{abstract}

We investigate the dissipative spectral form factor (DSFF)--a widely used probe of non-Hermitian quantum chaos--in the elliptic Ginibre unitary ensemble (eGinUE), which interpolates between the non-Hermitian Ginibre unitary ensemble (GinUE) and the Hermitian Gaussian unitary ensemble (GUE) via a symmetry breaking parameter. We derive exact finite-dimensional results and large-dimensional approximations for the DSFF, revealing a scaling relationship that connects the DSFF of eGinUE to that of GinUE and the spectral form factor of GUE. This relation explains the distinct time scales underlying the characteristic \emph{dip-ramp-plateau} structure across GinUE, GUE, and crossover regimes. Additionally, we refine estimates of dip-ramp and ramp-plateau transition times for different symmetry regimes. We validate our results with Monte Carlo simulations and demonstrate applications to paradigmatic quantum-chaotic systems: the crossover Sachdev-Ye-Kitaev model and the crossover Power-law Banded random matrices. We highlight an analogy between eGinUE eigenvalues and the positions of a rotating fermionic gas in a two-dimensional anisotropic trap.
\end{abstract}

\maketitle
\textit{Introduction.--}
The spectral form factor (SFF)~\cite{mehta2004random,haake2010quantum,forrester2010log} is a powerful probe of integrability and chaos, with applications ranging from quantum spin chains~\cite{vsuntajs2020quantum,PhysRevLett.123.190602,prasad2023long} and disordered systems~\cite{PhysRevLett.75.4389,DITTRICH1996267,vasilyev2020monitoring,prakash2021universal} to topological phases~\cite{sarkar2024spectral,okuyama2023spectral}. In its basic form, SFF is defined as the Fourier transform of the two-level correlation function with respect to the energy difference. Although the idea of using the Fourier transform to analyze spectral correlations is relatively old~\cite{berry1985semiclassical,leviandier1986fourier,sieber2001correlations,sieber2002leading,PhysRevLett.93.014103,PhysRevE.72.046207}, the SFF has recently gained prominence for the insights it provides into many-body physics~\cite{cotler2017black,PhysRevX.8.021062,PhysRevX.8.041019,PhysRevLett.121.264101} and for being experimentally accessible on quantum processors~\cite{PhysRevLett.134.010402}.
For random matrices, the SFF exhibits a \emph{dip-ramp-plateau} shape, serving as a diagnostic for detecting chaos~\cite{brezin1997spectral,gharibyan2018onset,saad2018semiclassical,gaikwad2019spectral,PhysRevLett.125.250602,forrester2021quantifying,charamis2025quenched}. In particular, \emph{ramp} beyond Thouless time ($T_\mathrm{Th}$)~\cite{schiulaz2019thouless,lezama2021equilibration} signals the emergence of random matrix behavior, which eventually levels off into the plateau at Heisenberg time ($T_\mathrm{H}$)~\cite{schiulaz2019thouless,lezama2021equilibration}. With the recent surge of interest in non-Hermitian physics~\cite{PhysRevLett.61.1899,PhysRevLett.79.1797,muller2012engineered,can2019random,sa2020spectral,PhysRevLett.100.103904,PhysRevLett.101.080402,ruter2010observation,RevModPhys.88.035002,feng2017non,el2018non,gupta2020parity,PhysRevB.101.014202,PhysRevB.106.L121102,ghatak2020observation,PhysRevB.98.165148,PhysRevB.98.085116,PhysRevLett.121.136802} and non-Hermitian random matrix theory (RMT)~\cite{ginibre1965statistical,PhysRevLett.60.1895,PhysRevLett.67.941, PhysRevE.108.054210, PhysRevE.111.034301}, analogous diagnostics are needed for probing non-Hermitian systems, including open systems. In addition to generalization of short-range correlations~\cite{PhysRevLett.123.090603,PhysRevLett.123.254101,PhysRevX.10.021019}, the dissipative spectral form factor (DSFF) extends the SFF to the complex plane, capturing universal features of dissipative chaotic systems~\cite{ghosh2022spectral,shivam2023many}, with exact RMT results available for both chaotic and integrable regimes~\cite{PhysRevLett.127.170602,garcia2023universality}.  This long-range spectral measure also displays the characteristic \emph{dip–ramp–plateau} behaviour on the absolute complex-time scale~\cite{PhysRevB.110.134316,li2024spectral,cipolloni2024dissipative}. In parallel, recent advances in experimental quantum platforms have enabled the measurement of spectral correlations~\cite{PhysRevLett.134.010402,wold2025experimental}, paving the way to realise DSFF-like probes experimentally.

In this work, we derive exact finite and large dimensional expressions for the DSFF of the elliptic Ginibre unitary ensemble (eGinUE)~\cite{fyodorov1997almostcrossover,fyodorov1997almosteigenvaluedensity,khoruzhenko2015nonhermitian,akemann2003characteristic}, which interpolates between the Ginibre unitary ensemble (GinUE)~\cite{ginibre1965statistical} and the Gaussian unitary ensemble (GUE)~\cite{forrester2010log,mehta2004random} via a symmetry breaking parameter. A key result is an exact ``scaling" relation linking the DSFF of eGinUE to the SFF of GUE, valid for all dimensions and symmetry-breaking strengths. This mapping allows the known SFF results for GUE to be directly applicable to DSFF for eGinUE and GinUE. From large-dimensional analysis, we provide refined estimates of $T_\mathrm{Th}$ and $T_\mathrm{H}$ that depend on the degree of hermiticity breaking. The universality of these results is demonstrated through their ability to capture the late-time DSFF of the crossover Sachdev-Ye-Kitaev (cSYK) and crossover Power-law banded random matrix model (cPLBRM). We further highlight a mapping between the eigenvalues of eGinUE and the positions of a rotating fermionic gas in its unique ground state within a two-dimensional anisotropic trap~\cite{akdeniz2005shell,ghazanfari2010rapidly,medjedel2019exact}.

\textit{Dissipative Spectral Form Factor (DSFF).--}
For an ensemble of real eigenvalues $\{E_i\}_{i=1}^{D}$, the ensemble-averaged SFF is
\begin{align}{\label{SFFbasic}} 
\mathrm{SFF}(t)= \left\langle \left|\sum_{n=1}^{D}e^{it E_{n}}\right|^2\right\rangle =\left \langle \sum_{n,m=1}^{D}e^{it(E_{n}-E_{m})} \right \rangle,
\end{align}
where `$t$' is the \emph{time} variable and $\langle{\cdot\rangle}$ represents the ensemble average. 
Analogous to the SFF, the DSFF is defined for complex spectra $\{z_i\}_{i=1}^{D}$,
where $z_{i}=x_{i}+iy_{i}; x_{i}, y_{i}\in \mathbb{R}$, corresponding to a generic $D$-dimensional non-Hermitian matrix. The ensemble-averaged and scaled DSFF is then given by~\cite{PhysRevLett.127.170602}, 
\begin{align}\label{DSFFdef_main}
    \mathcal{F}(t,s)=1+\frac{1}{D}\left\langle \sum_{n,m=1 \atop (n\ne m)}^{D}e^{it(x_{n}-x_{m})+is(y_{n}-y_{m})} \right\rangle.
\end{align}
Here, $t$ and $s$ are two generalized time variables conjugate to the real $(x_{n}-x_{m})$ and imaginary $(y_{n}-y_{m})$ parts of the difference between two complex eigenvalues, $z_{n}$ and $z_{m}$. When the spectrum is real, the SFF($t$) in Eq.~\eqref{SFFbasic} is recovered from the DSFF in Eq.~\eqref{DSFFdef_main} as a function of $t$ for all finite $s$. More generally, the complex time variable $T=t+is\equiv|T|e^{i\varphi}$ can be introduced to analyze DSFF in the complex plane~\cite{PhysRevLett.127.170602}. It was established in Ref.~\cite{PhysRevLett.127.170602}, akin to SFF in closed systems, the DSFF exhibits a \emph{dip-ramp-plateau} behavior as a function of $|T|$ for standard non-Hermitian Gaussian ensembles, namely, the Ginibre ensembles. The nature of the ramp, however, in this case is nonlinear. The exact DSFF for the two-dimensional Poissonian case is also provided in Ref.~\cite{PhysRevLett.127.170602}.

\textit{Elliptic Ginibre Unitary Ensemble (eGinUE).--} 
The eGinUE interpolates between Hermitian random matrices governed by Wigner–Dyson statistics~\cite{fyodorov1997almostcrossover} and non-Hermitian Ginibre ensembles characterized by complex spectra~\cite{ginibre1965statistical}. A symmetry-breaking parameter drives the crossover, deforming the spectrum from the complex plane (non-Hermitian) to the real line (Hermitian) and altering spectral correlations. Physically, eGinUE can be interpreted as a two-dimensional one-component plasma subjected to a quadrupolar field~\cite{francesco1994laughlin,forrester1996two,forrester2010log}, extending Dyson’s log-gas analogy for eigenvalues in Hermitian random matrix ensembles to the non-Hermitian one. A eGinUE matrix can be decomposed in terms of its Hermitian and skew-Hermitian components~\cite{fyodorov1997almostcrossover}, viz., $\textbf{H}=\textbf{S}+i \textbf{A},$ where $\textbf{S}$ and $\textbf{A}$ are $D$-dimensional statistically independent Hermitian matrices chosen from GUE with probability densities given in the Supplemental Material~\cite{supp}. Using their distributions, the probability density of $\bH$ is~\cite{fyodorov1997almostcrossover}  
\begin{align}{\label{GinUEMM2_main}}
    \mathcal{P}_{H}(\bH)= \left(\frac{D}{\pi v^{2}\sqrt{1-\tau^{2}}}\right)^{D^2} 
    e^{-\frac{D\,\tr(\bH\bH^{\dagger}-\tau \,\mathrm{Re}(\bH^{2}))}{(1-\tau^{2})v^2}},
\end{align}
where $\mathrm{Re}$ denotes the real part. The degree of this symmetry-breaking is regulated by the crossover parameter $\tau$, and the additional parameter $v^2>0$ fixes the scale of the problem~\cite{supp}. In the limits $\tau\to 0$ and $\tau\to 1$, we obtain GinUE and GUE, respectively~\cite{supp}. The term ``elliptic'' refers to the fact that, as $D\!\to\!\infty$, the eigenvalues $\{z_i\}$ of $\bH$ lie within an ellipse~\cite{girko1986elliptic,PhysRevLett.60.1895},  $\frac{x^{2}}{(1+\tau)^{2}}+\frac{y^{2}}{(1-\tau)^{2}}\le v^{2}$. For $\tau=0$, they are distributed uniformly over a disk of radius $v$, while $\tau\!\to\!1$ collapses them to the real line. We set $v=1$ hereafter for simplicity.

\textit{Calculations and results.--} 
We present exact finite-dimensional expressions for the DSFF of eGinUE together with their large-dimensional approximations. This calculation involves the joint probability density $\mathcal{P}_{D}(\{z_{i}\}_{i=1}^{D})$ of the complex eigenvalues, known explicitly for the matrix ensemble given in Eq.~\eqref{GinUEMM2_main} from~\cite{fyodorov1997almostcrossover}. The DSFF in Eq.~\eqref{DSFFdef_main} can be computed using the correlation kernel~\cite{supp}, yielding
\begin{align}
\label{DSFFmain_main}
 \mathcal{F}(t,s)=1+\mathcal{F}_{\text{dis}}(t,s)-\mathcal{F}_{\text{conn}}(t,s), 
\end{align}
where, $1$ is the contribution from the contact part. The disconnected part, $\mathcal{F}_{\text{dis}}$, of the DSFF is given by
\begin{align}{\label{Fdis_main}}
  \mathcal{F}_{\text{dis}}(t,s)=\frac{1}{D}e^{-\frac{(1+\tau)t^{2}}{2\,D}-\frac{(1-\tau)s^{2}}{2\,D}}\big[L_{D-1}^{(1)}(\mathbb{T})\big]^{2},
\end{align}
and the connected part, $\mathcal{F}_\text{conn}$, reads
\begin{align}{\label{Fconn_main}}
\nonumber
\cF_{\text{conn}}(t,s)&=\frac{1}{D}e^{-\frac{(1+\tau)t^{2}}{2\,D}-\frac{(1-\tau)s^{2}}{2\,D}}\\
&~~~~~~~~\times \sum_{n,m=0}^{D-1}\frac{m!}{n!}\mathbb{T}^{n-m}\big[L_m^{(n-m)}(\mathbb{T})\big]^{2}.
\end{align}
In Eqs.~\eqref{Fdis_main} and ~\eqref{Fconn_main}, $L_a^{(k)}(u)$ is the associated Laguerre polynomial and we have defined,
$\mathbb{T}=\frac{(1+\tau)^{2}t^{2}+(1-\tau)^{2}s^{2} }{4\,D}=\frac{\eta^2}{D}|T|^2,$ with $\eta=(1/2)[(1+\tau)^{2}\cos^2\varphi+(1-\tau)^{2}\sin^2\varphi]^{1/2}$, where we recall that $T=t+is$. The detailed derivation of Eqs.~\eqref{Fdis_main} and~\eqref{Fconn_main} is presented in the Supplemental Material~\cite{supp}.
At early times $|T|\lesssim T_\mathrm{Th}$, the DSFF captures the energy level correlations at scales significantly larger than the mean energy level spacing of the system, exhibiting a decay behavior, referred to as the \textit{dip}. The oscillations around the \emph{dip} is dominated by the $\mathcal{F}_{\text{dis}}$. In the intermediate regime $T_\mathrm{Th}\lesssim |T| \lesssim T_\mathrm{H}$, the connected part dominates, resulting in a \textit{ramp}-like behavior, crucial for identifying quantum chaos. Beyond $T_\mathrm{H}$, the DSFF tends to saturate to a constant \textit{plateau} value (of unity) governed by the contact term.
It should be noted that the DSFF in this case depends on the combination of $t^2=|T|^2\cos^2\varphi$ and $s^2=|T|^2\sin^2\varphi$, which makes it invariant under the transformation $\varphi\to \pi\pm \varphi$, where $\varphi$ is the argument of $T$. Equations~\eqref{Fdis_main} and~\eqref{Fconn_main} constitute the first key contribution of our paper. When one of the limiting cases, $\tau=0$, is considered, Eq.~\eqref{DSFFmain_main} leads to the GinUE result~\cite{PhysRevLett.127.170602,garcia2023universality} (see also the Supplemental Material~\cite{supp})
\begin{align}
\label{Fginue_main}
\nonumber
&\mathcal{F}^\mathrm{GinUE}(T)= 1+\frac{e^{-\frac{|T|^{2}}{2\,D}}}{D}\left[   L_{D-1}^{(1)}\bigg(\frac{|T|^2}{4D}\bigg)\right]^{2}\\
&-\frac{e^{\frac{|T|^{2}}{2\,D}}}{D}\sum_{n,m=0}^{D-1}\frac{m!}{n!}\bigg(\frac{|T|^{2}}{4D}\bigg)^{n-m}\left[L_m^{(n-m)}\bigg(\frac{|T|^{2}}{4\,D}\bigg)\right]^{2}.
\end{align}
Setting $\tau=1$, we obtain the GUE limit~\cite{del2017scrambling} 
\begin{align}\label{Fgue_main}
\nonumber
& \mathcal{F}^\mathrm{GUE}(t)= 1+\frac{e^{-\frac{t^{2}}{D}}}{D}\left[    L_{D-1}^{(1)}  \bigg(\frac{t^2}{D}\bigg)   \right]^{2}\\
&-\frac{e^{-\frac{t^{2}}{D}}}{D}\sum_{n,m=0}^{D-1}\frac{m!}{n!}\left(\frac{t^{2}}{D}\right)^{n-m}\left[L_m^{(n-m)}\bigg(\frac{t^{2}}{D}\bigg)\right]^{2}.
\end{align}  Eq.~\eqref{Fgue_main} depends on $\mathrm{Re}(T)=t=|T|\cos\varphi$, just serving as the scaling of the time variable as far as GUE is concerned. 

\begin{figure}
\includegraphics[width=0.85\columnwidth]{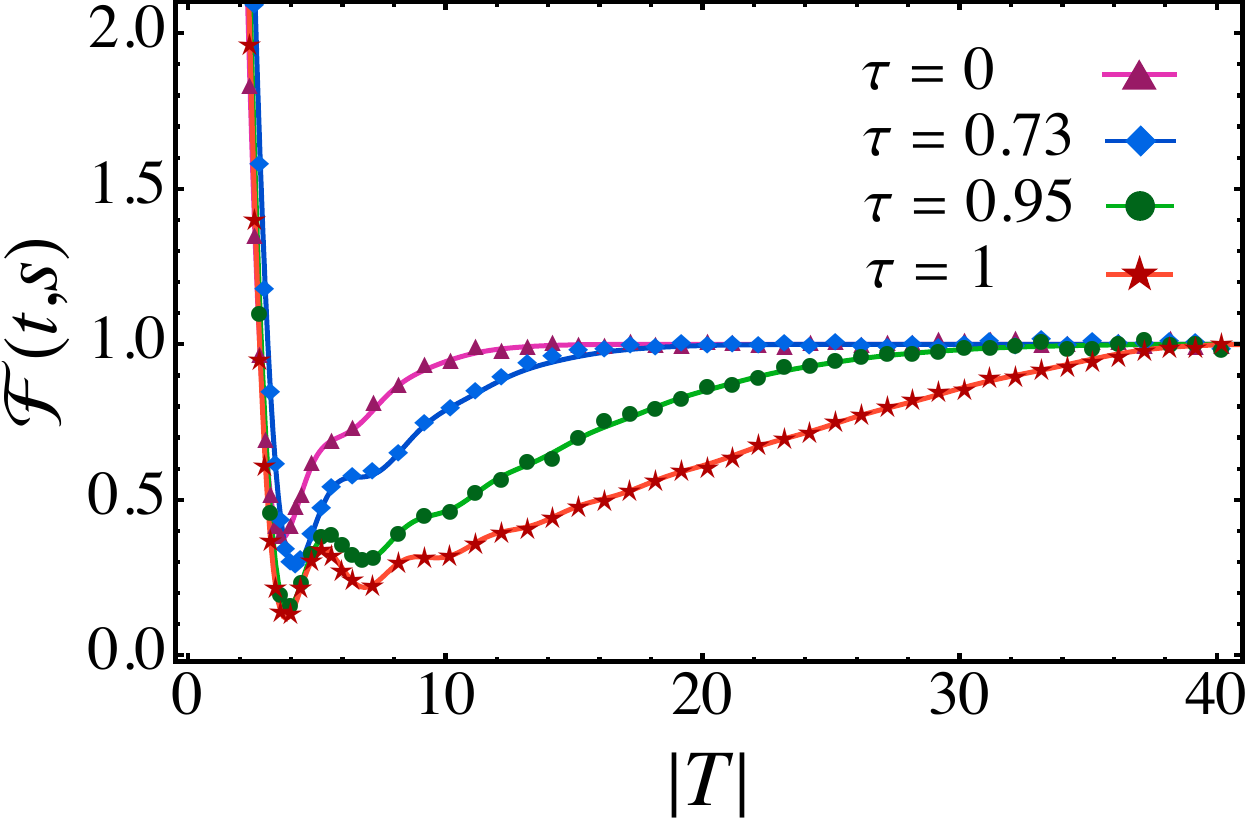}
\caption{Plots of the eGinUE DSFF $\mathcal{F}(t,s)$ as a function of $|T|$ for $D=10$ and $\varphi=\pi/3$. The solid curves represent exact analytical results, Eq.~\eqref{DSFFmain_main}, while symbols are based on Monte Carlo simulations of the eGinUE matrix model. The crossover from GinUE to GUE symmetry is clearly observed with varying symmetry-breaking parameter $\tau$.}
\label{fig1}
\end{figure}

Figure~\ref{fig1} shows the eGinUE DSFF for $D=10$, $\varphi=\pi/3$, comparing the exact result of Eq.~\eqref{DSFFmain_main} (solid) with the numerical data obtained from the Monte Carlo simulations of the eGinUE matrix model of $10^4$ matrices (symbols). The agreement is excellent and displays the characteristic \emph{dip–ramp–plateau}.
A direct comparison of Eqs.~\eqref{Fginue_main} and~\eqref{Fgue_main} reveals an intriguing ``scaling relation" between their disconnected (second term) and connected (third term) parts, viz.
\begin{align}
\label{FGinGUscale_main}
\mathcal{F}^\mathrm{GinUE}_\mathrm{dis,conn}(T)=e^{\frac{-|T|^2}{4D}}\mathcal{F}^\mathrm{GUE}_\mathrm{dis,conn}(|T|/2).
\end{align}
 In fact, the DSFF for the eGinUE [Eq.~\eqref{DSFFdef_main}] can similarly be expressed in terms of the GUE~\eqref{Fgue_main} result as
\begin{equation}
\label{scaling_main}
\begin{aligned}
\cF_\mathrm{dis}(t,s)
  &= e^{-\frac{(1-\tau^2)(t^2+s^2)}{4D}}
\mathcal{F}^\mathrm{GUE}_\mathrm{dis}(\sqrt{D\mathbb{T}}),\\
\cF_\mathrm{conn}(t,s)
  &= e^{-\frac{(1-\tau^2)(t^2+s^2)}{4D}}
\mathcal{F}^\mathrm{GUE}_\mathrm{conn}(\sqrt{D\mathbb{T}}).
\end{aligned}
\end{equation}
Besides being aesthetically pleasing, the expressions in Eq.~\eqref{scaling_main} are immensely powerful. They reveal a deep connection among the different symmetry regimes of the eGinUE, including the GinUE and GUE limits, and enable efficient derivations of large-$D$ approximations for the DSFF.

We now present large-$D$ approximations for the DSFF, deduced in detail in the Supplemental Material~\cite{supp}. Starting from Eq.~\eqref{DSFFmain_main} we find the disconnected and connected parts of the DSFF of eGinUE in this limit are, respectively
\begin{align}
\label{FdisAsy_main}
\mathcal{F}_\mathrm{dis}(t,s)&\approx\frac{1}{\mathbb{T}}\,e^{-\frac{(1-\tau^2)(t^2+s^2)}{4D}}J_1\big(2\sqrt{D\mathbb{T}}\big)^2,
\end{align}
\begin{align}
\label{FconnAsy_main}
\mathcal{F}_\mathrm{conn}(t,s)\approx e^{-\frac{(1-\tau^2)(t^2+s^2)}{4D}}R\big(\sqrt{D\mathbb{T}}\big).
\end{align}
Here, $J_{\alpha}(u)$ is the Bessel function of order $\alpha$, and $R(t)=\frac{2}{\pi}\csc^{-1}\left(\frac{2D}{\sqrt{4D^2-t^2}}\right)-\frac{t \sqrt{4D^2-t^2}}{2\pi D^2},$ for $t< 2D$, while $R(t)=0$ for $t\ge 2D$~\cite{liu2018spectral}. Equations~\eqref{FdisAsy_main} and~\eqref{FconnAsy_main} can be utilized to reveal several interesting details about Thouless time $T_\mathrm{Th}$, Heisenberg time $T_\mathrm{H}$, and the behavior of the dip and ramp regions in the large-$D$ limit. 

\begin{figure*}
\includegraphics[width=0.98\linewidth]{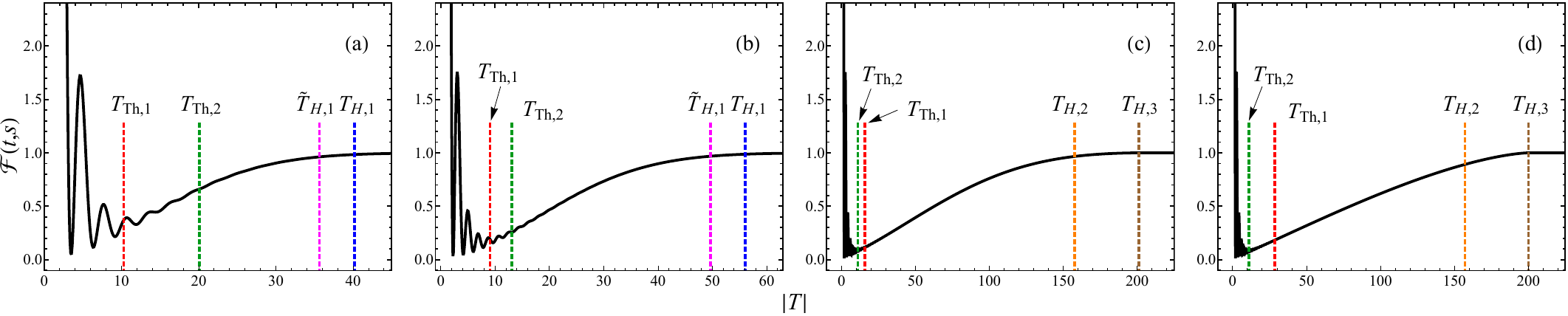}
\caption{Plots of DSFF, using $1+\mathcal{F}_\mathrm{dis}-\mathcal{F}_\mathrm{conn}$, where $\mathcal{F}_\mathrm{dis}$ and $\mathcal{F}_\mathrm{conn}$ are given by Eqs.~\eqref{FdisAsy_main} and~\eqref{FconnAsy_main}, respectively for $D=100$, $\varphi=0$ and varying $\tau$: (a) $\tau=0.1$, (b) $\tau=0.7$, (c) $\tau=0.99$, and (d) $\tau=0.9995$. The theoretical estimates for Thouless time \big($T_\mathrm{Th,1}\approx(\frac{6}{\pi \eta^3 (1-\tau^2)})^{1/5} D^{2/5}$, $T_\mathrm{Th,2}\approx(\frac{3}{2})^{1/4}\frac{D^{1/2}}{\eta}$\big), and Heisenberg time \big($T_{\mathrm{H},1}\approx4\sqrt{D/(1-\tau^2)}$, $\widetilde{T}_{\mathrm{H},1}
\approx2\sqrt{\pi}\sqrt{D/(1-\tau^2)}$, $T_{\mathrm{H},2}\approx\pi D/(2\eta)$, $T_{\mathrm{H},3}\approx2D/\eta$\big) are shown as vertical lines (see Table~\ref{tab:TH_definitions} in~\cite{supp}). For (a) and (b), where $\tau<\tau_c$, $T_\mathrm{Th,1}$ serves as a better estimate for Thouless time, whereas for (c), (d) for which $\tau\gtrsim \tau_{c}$ the value $T_\mathrm{Th,2}$ serves better. Similarly, for the Heisenberg time $T_\mathrm{H,1}$ and $\widetilde{T}_{\mathrm{H},1}$  serves as a good estimate for (a) and (b). On the other hand, for (c) and (d), $T_\mathrm{H,2}$ and $T_\mathrm{H,3}$ serve as better estimates, respectively. $T_\mathrm{H,1}$ and $\widetilde{T}_{\mathrm{H},1}$
 values in panels (c) and (d) lie beyond the time scale shown.}
\label{fig3}
\end{figure*}

For time values in the vicinity of $T_{\mathrm{H}}$, the contribution due to the connected part becomes small in comparison to the contact part (unity), which results in the plateau. Moreover, the disconnected part becomes negligible at a much earlier time. Thus, $T_{\mathrm{H}}$ may be estimated using $\cF_\mathrm{conn} \approx0$. Expanding Eq.~\eqref{FconnAsy_main} up to $\mathcal{O}(1/D)$ and solving for time leads to $T_{\mathrm{H}}\sim\mathcal{O}(\sqrt{D})$ for $\tau=0$ (GinUE), and $T_{\mathrm{H}}\sim\mathcal{O}(D)$ for $\tau=1$ (GUE), as is known from earlier works~\cite{PhysRevLett.127.170602,liu2018spectral}. However, the numerical estimate for $T_{\mathrm{H}}$ obtained in this manner does not compare very well with the simulation data for large but finite $D$. To address this, we define \( \tau_c := 1 - c/D \), with \( c \sim \mathcal{O}(1) \). For finite $D$, $\tau < \tau_c$ corresponds to the strongly non-Hermitian side, $\tau > \tau_c$ lies on the strongly Hermitian side, and $\tau \approx \tau_c$ marks the crossover window between the two. For $\tau <\tau_c$, the exponential term in Eq.~\eqref{FconnAsy_main} leads to a much faster decay of $\cF_\mathrm{conn}$ towards 0 compared to the factor $R(\sqrt{D\mathbb{T}})$ that remains $\mathcal{O}(1)$. To obtain a reasonable estimate for $T_{\mathrm{H}}$ in this regime, we may consider $\cF_{\mathrm{conn}}=\exp\big(-\frac{(1-\tau^2)(t^2+s^2)}{4D}\big)=\exp\big(-\frac{(1-\tau^2)|T|^2}{4D}\big)=e^{-4}\approx 0.018$ as a convenient choice, since directly setting $\mathcal{F}_{\mathrm{conn}} = 0$ would otherwise 
result in a divergent $T_{\mathrm{H}}$. This leads to $T_\text{H} \approx 4\sqrt{D/(1 - \tau^2)}$. While obtained from a heuristic scaling choice, this expression provides a useful approximation. Alternatively, a more concrete definition of $T_{\text{H}}$ is given in terms of the mean level spacing $\delta$ between adjacent eigenvalues, with $T_{\text{H}} = 2\pi/\delta$~\cite{haake2010quantum}. Expressing $\delta$ through the eigenvalue density in the strong non-hermiticity regime~\cite{fyodorov1998universality} gives $T_{\text{H}} \approx 2\sqrt{\pi}\sqrt{D/(1 - \tau^2)}$ offering a compact analytical estimate. For $\tau\approx\tau_c$, by equating the expansion of $\cF_\mathrm{conn}$ up to $\mathcal{O}(1/D)$ to zero, we obtain $T_\mathrm{H}\approx \pi D/(2\eta)$. This value is less than $2D/\eta$ beyond which $\cF_\mathrm{conn}$ is identically equal to zero due to the $R(\sqrt{D\mathbb{T}})$ factor. For $\tau_{c}< \tau \leq 1$, $T_{\mathrm{H}}\approx2D/\eta$ serves as a better and natural estimate. To estimate $T_\mathrm{Th}$, we determine the minimum of the local DSFF envelope curve in the \emph{dip} region using the leading behavior of $\mathcal{F}_\mathrm{conn}$ [Eq.~\eqref{FconnAsy_main}] and the nonoscillatory part of $\mathcal{F}_\mathrm{dis}$ [Eq.~\eqref{FdisAsy_main}] in large-$D$ expansion. For $\tau < \tau_c$, the estimate $T_\mathrm{Th} \approx \left( \frac{6}{\pi \eta^3 (1 - \tau^2)} \right)^{1/5} D^{2/5}$ performs very well, whereas for $\tau \gtrsim \tau_c$, we obtain $T_\mathrm{Th} \approx \left( \frac{3}{2} \right)^{1/4} \frac{D^{1/2}}{\eta}$. These yield the scalings $T_\mathrm{Th}\sim\mathcal{O}(D^{2/5})$ and $\mathcal{O}(D^{1/2})$ for the GinUE and GUE limits, respectively, consistent with earlier findings~\cite{PhysRevLett.127.170602,del2017scrambling}.This approximation, however, breaks down in the limit \(\varphi \to \pi/2\), since in this case \(1/\eta \to \infty\), where \(\varphi = \tan^{-1}(s/t)\).
The details of these estimates for $T_\mathrm{H}$ and $T_\mathrm{Th}$ are provided in the Supplemental Material~\cite{supp}.

In Fig.~\ref{fig3}, we test the estimates for $T_\mathrm{Th}$ and $T_\mathrm{H}$ for $D=100$, $\varphi=0$ and several values of $\tau$. The DSFF curves agree with the estimated times, and similar consistency is found for other $(D,\varphi)$ across the $\tau$-regimes. Having established the validity of our analytical findings within eGinUE, we next examine their relevance to physical models by studying crossover variants of the SYK and PLBRM models.

\begin{figure}
\includegraphics[width=0.96
\columnwidth]{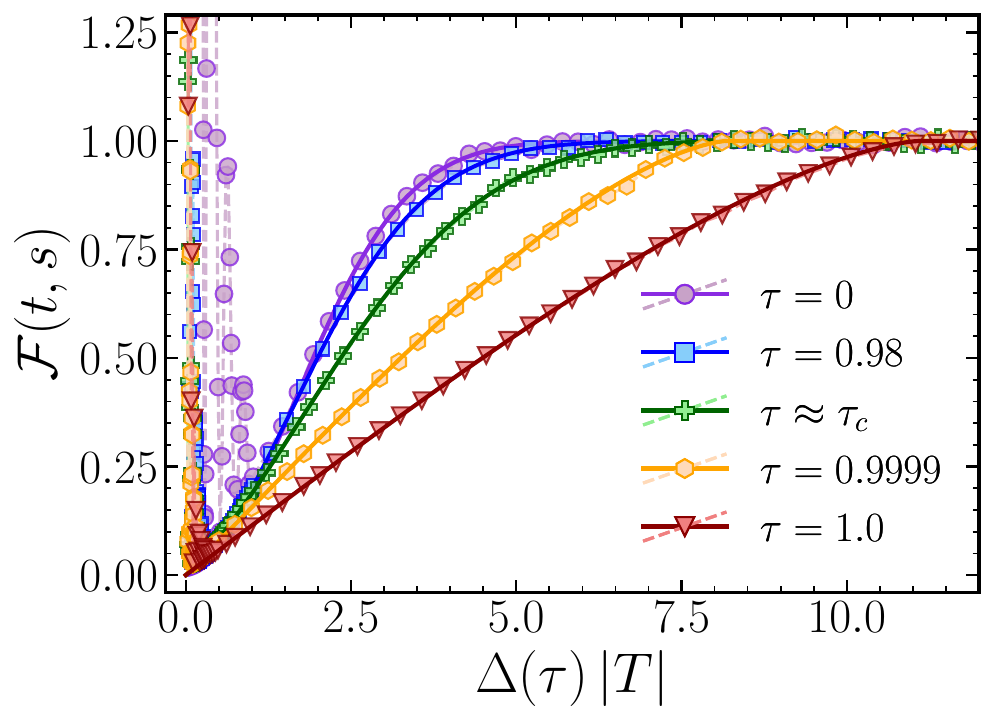}
\caption{The DSFF of the $q = 4$ cSYK model (symbols), given in Eq.~\eqref{cSYK}, for $N = 18$, normalized by $2^{N/2-1}$ non-degenerate eigenvalues and averaged over $5 \times 10^4$ matrices. We compare it with the large-$D$ asymptotic eGinUE result (solid curves), $1 - \mathcal{F}_\mathrm{conn}$ [Eq.~\eqref{FconnAsy_main}], evaluated with $D = 2^8$ at $\varphi = \pi/4$ and find excellent agreement. Varying $\tau$ demonstrates the GinUE–GUE crossover, with $\tau_c = 1 - 1/2^8$.
}
\label{figsyk}
\end{figure}

\textit{Crossover SYK (cSYK).--} 
Originally introduced in the study of spin liquids~\cite{sachdev1993gapless}, the SYK model~\cite{kitaev2015simple,RevModPhys.94.035004} has since become a central framework in the study of many-body quantum chaos~\cite{PhysRevLett.117.111601,PhysRevD.94.106002}.
It has gained renewed interest through connections to holography~\cite{PhysRevLett.105.151602,PhysRevD.106.046002}, high-temperature superconductivity~\cite{cai2021superconducting,PhysRevResearch.3.033117}, and strange metallic transport~\cite{PhysRevX.5.041025,PhysRevResearch.2.023366}, among others. We introduce the crossover SYK (cSYK) model as a testbed to validate our analytical results. The model Hamiltonian represents $N$ strongly interacting Majorana fermions
$(\chi_i = \chi_i^\dagger)$, obeying $\{\chi_i, \chi_j\} = 2\delta_{ij}$, 
coupled via infinite-range $q$-body interactions
\begin{align}
\label{cSYK}
\textbf{H}^{\mathrm{cSYK}}=\sum_{i_{1}<i_{2}<\ldots<i_{q}}^{N}(J_{i_{1}i_{2}\ldots i_{q}}+iM_{i_{1}i_{2}\ldots i_{q}})\chi_{i_{1}}\chi_{i_{2}}\cdots\chi_{i_{q}}.
\end{align}
The couplings $J_{i_{1}i_{2}\ldots i_{q}}$ and $M_{i_{1}i_{2}\ldots i_{q}}$ are real random variables chosen from normal distributions $\mathcal{N}\left(0,\sqrt{(1+\tau)\frac{(q-1)!}{N^{q-1}}}\right)$ and $\mathcal{N}\left(0,\sqrt{(1-\tau)\frac{(q-1)!}{N^{q-1}}}\right)$, respectively. In the limits $\tau\rightarrow 0$ and $\tau\rightarrow1$, one obtains the conventional limits of non-Hermitian~\cite{PhysRevX.12.021040,PhysRevD.107.066007} and Hermitian~\cite{PhysRevD.94.126010,cotler2017black,PhysRevB.95.115150} SYK, respectively, which are well studied in the context of RMT. Figure~\ref{figsyk} shows results of DSFF of the cSYK model for different values of $\tau$ and its comparison to large-$D$ analytical results of eGinUE. Both are plotted against the rescaled time variable $\Delta(\tau)|T|$, for direct comparison. Here, $\Delta(\tau)$ denotes the mean level spacing, obtained separately from numerical simulations of the respective spectra. Since the SFF (and its dissipative counterpart) is not self-averaging~\cite{PhysRevLett.78.2280}, we average over a large number of realizations to obtain a smooth curve for a large but finite $N$. However, if $N$ is too large, $\tau_c \to 1$ as $1-e^{-\lambda N}$ with $\lambda=(1/2)\ln2$, thus drastically narrowing the $\tau\gtrsim\tau_c$ regime. Simulations for larger system sizes have also been performed and are presented in the Supplemental Material~\cite{supp}.

\begin{figure}
\includegraphics[width=0.96\columnwidth]{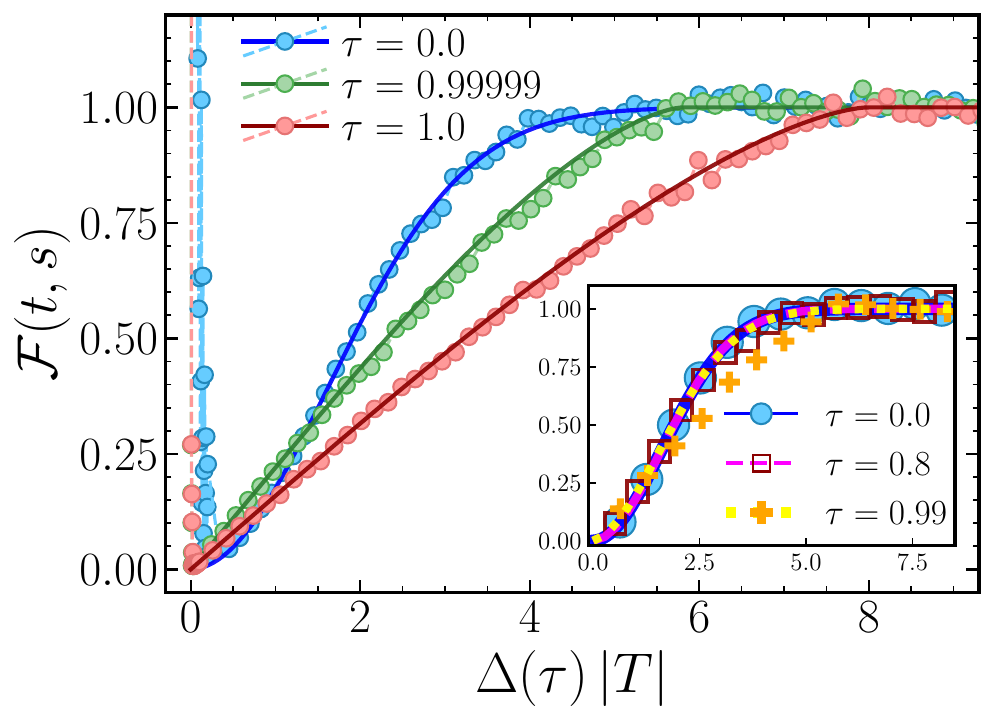}
\caption{The DSFF of the cPLBRM model (symbols) given in Eq.~\eqref{cPLBRM} for matrix size $4096$, averaged over $4000$ realizations using the full eigenspectra. This is compared with the asymptotic eGinUE DSFF results (solid curves), $1 - \mathcal{F}_\mathrm{conn}$ [Eq.~\eqref{FconnAsy_main}], evaluated with $D = 4096$. We set $W = 3$ and $p = 0.5$ to place the system well within the chaotic regime, with $b = 1$ fixed for simplicity. Varying $\tau$ demonstrates the GinUE–GUE crossover. Throughout, we set $\varphi = 0$. The inset shows the same comparison for additional values of $\tau$, where analytical results are shown by solid, dashed and dotted lines for the three values of $\tau$.
}
\label{figplbrm}
\end{figure}

\textit{Crossover PLBRM (cPLBRM).--} Another paradigmatic model is the Power-law Banded Random Matrix (PLBRM)~\cite{levitov1989absence,PhysRevLett.64.547,PhysRevE.54.3221,RevModPhys.80.1355}, widely used to study universal features of disordered quantum systems. It captures parameter-dependent localization transitions, from delocalized to localized phases, through eigenvector and spectral correlations~\cite{PhysRevE.54.3221,PhysRevB.62.7920,PhysRevLett.84.3690}. Here, we introduce the crossover PLBRM (cPLBRM) and study the effect of Hermiticity breaking on complex systems through its DSFF.
The ensemble of such matrices is given by
\begin{align}
\label{cPLBRM}
\textbf{H}_{ij}^{\mathrm{cPLBRM}}=\epsilon_{i}\,\delta_{ij}+g_{ij},
\end{align}
where $\epsilon_{i}$ and $g_{ij}=g_{ji}^*$ are complex random variables selected from the zero-mean uniform distributions with bounds
\begin{align}
\left|\mathrm{Re}(\epsilon_{i})\right|\leq\frac{(1+\tau)}{2}W,\,
\left|\mathrm{Im}(\epsilon_{i})\right|\leq\frac{(1-\tau)}{2}W,
\label{cPLBRM_epsilon}\\
\left|\mathrm{Re}(g_{ij})\right|,\left|\mathrm{Im}(g_{ij})\right|\leq [2(|i-j|^{2}+b^{2})]^{-p/2}.
\label{cPLBRM_g}
\end{align}
Here $W$ is the onsite disorder strength. The off-diagonal terms decay as a power law with exponent $p$ and $b$ controls the bandwidth of the decay. For $\tau = 1$, the model reduces to the Hermitian PLBRM~\cite{PhysRevE.54.3221,PhysRevB.62.7920,PhysRevLett.84.3690} and corresponds to GUE statistics in its chaotic regime. In the $\tau<1$ regime, the model is non-Hermitian due to the presence of complex diagonal elements, with the $\tau = 0$ case yielding GinUE statistics~\cite{de2023non,ghosh2023eigenvector}. As shown in Fig.~\ref{figplbrm}, the DSFF for cPLBRM exhibits clear signatures of GinUE–GUE crossover, showing good agreement with asymptotic eGinUE expressions. Here, the time axis is also rescaled using the mean level spacing $\Delta(\tau)$, extracted numerically. Nevertheless, unlike the cSYK case discussed earlier, this model shows strong finite-size effects in the crossover regime, governed by $\tau$~\cite{supp}. Using large matrices reduces these artifacts, but also narrows the crossover window and drives the system toward one of the two asymptotic symmetry classes~\cite{supp}. This trend is visible in the inset of Fig.~\ref{figplbrm}, where the DSFF curves for several $\tau$ values in the non-Hermitian regime cluster closely around the GinUE result. However, for a large but finite system, $\tau\approx1$ (e.g. $\tau = 0.99999$ here) can still remain in the crossover rather than overlapping with GUE. The inset further shows a mismatch (a finite-size effect) for $\tau = 0.99$, a model-specific deviation that diminishes as $\tau$ approaches the Hermitian limit, as seen for $\tau = 0.99999$ in the main plot~\cite{supp}. Together, these observations highlight the delicate balance between system size and non-Hermiticity required to reliably capture the DSFF crossover behaviour in cPLBRM. 

\textit{Discussion.--} Non-interacting fermions in confining traps form determinantal point processes closely linked to random matrix ensembles~\cite{PhysRevLett.112.254101,PhysRevLett.114.110402,dean2015universal,PhysRevA.94.063622,PhysRevA.99.021602,dean2019noninteracting}. For instance, at $T=0$, fermions in a two-dimensional rapidly rotating isotropic harmonic trap map onto the GinUE, where the many-body ground state forms a uniform circular droplet~\cite{PhysRevA.99.021602,smith2022,msg2021}. Introducing a small anisotropy into the trap transforms this droplet into constant density elliptical plateaus that match the average eigenvalue density of eGinUE in its unique ground state~\cite{ghazanfari2010rapidly, medjedel2019exact}. Thus, in suitably prepared cold-atom systems, fermion position correlations~\cite{PhysRevLett.134.183403,PhysRevLett.134.183402} can be mapped to eGinUE eigenvalue correlations, suggesting that the DSFF \emph{ramp–plateau} structure could serve as an experimental signature of chaos.

We study the eGinUE and derive exact and asymptotic expressions for its DSFF. A scaling relation is established between the DSFF of eGinUE and the SFF of GUE, enabling the former to be obtained directly from the latter. Refined estimates of the $T_{\mathrm{H}}$ and $T_{\mathrm{Th}}$ are provided across the GinUE–GUE crossover. All results are validated through Monte Carlo simulations. The findings capture the DSFF behaviour of paradigmatic non-Hermitian systems such as the cSYK and cPLBRM models. We further point out an analogy between eGinUE eigenvalues and the positions of a rotating fermionic gas in a two-dimensional anisotropic trap, providing additional physical motivation.

The crossover behaviour explored here can be generalized to other spectral observables, enabling a broader understanding of universality and scaling in non-Hermitian ensembles. The present framework can also be generalized to real~\cite{forrester2008skew,byun2021real, forrester2024local,akemann2025probability} and symplectic~\cite{kanzieper2002eigenvalue,byun2023universal,byun2025three} eGinEs, thereby covering all Dyson symmetry classes. Finally, applying these insights to open quantum systems—such as dissipative spin chains~\cite{PhysRevE.108.054210,zhou2025diagnosting}, photonic lattices~\cite{PhysRevB.101.014202}, and cold-atom platforms~\cite{PhysRevA.99.021602,smith2022,msg2021}—could establish direct links between RMT and experimentally measurable observables in non-Hermitian many-body physics. 

\textit{Note added.--} While this manuscript was in preparation, one of the coauthors, Santosh Kumar, passed away. An article reviewing his scientific contributions has since appeared~\cite{forrester2025memoriam}.

\textit{Acknowledgments.--} 
S.S. acknowledges financial support from Shiv Nadar Institution of Eminence and thanks the hospitality of the International Centre for Theoretical Sciences, Bangalore. S.K. was supported by SERB, DST, Government of India via Project No. CRG/2022/001751. A.S. acknowledges support from the Canada First Research Excellence Fund through an Institut Quantique Postdoctoral Fellowship. We are deeply grateful to our late coauthor, Santosh Kumar, for his invaluable contributions to the conceptualization and development of this work. S.S. gratefully acknowledges R. Ghosh and S. Pandey for invaluable discussions.

\bibliography{references}


\setcounter{equation}{0}
\setcounter{figure}{0}
\renewcommand{\theequation}{S\arabic{equation}}
\renewcommand{\thefigure}{S\arabic{figure}}

\onecolumngrid
\newpage
\begin{center}
{\textbf{\underline{Supplemental Material}}}
\end{center}


This Supplementary Material is organized as follows. In Sec.\ref{AppDSFF}, we provide a thorough derivation of our exact analytical results for the DSFF of the eGinUE using rigorous mathematical calculations. In Sec.\ref{AppDSFFAsy}, we discuss the derivation of the large-dimensional results of the above. A detailed discussion of the Heisenberg time and Thouless time is presented in Sec.~\ref{AppTime}. Section~\ref{Application} provides additional discussions on the model-specific behaviours of the cSYK and cPLBRM models, along with a comparison of their DSFF characteristics against the asymptotic eGinUE results.

\section{
Exact Analytical expression of the DSFF for the \MakeLowercase{e}G\MakeLowercase{in}UE}\label{AppDSFF}
As discussed in the main text, the non-Hermitian random matrix belonging to eGinUE is defined as~\cite{fyodorov1997almostcrossover}
\begin{align}\label{GinUEMM1}  
\textbf{H} = \textbf{S} + i\textbf{A},  
\end{align}  
where $\textbf{S}$ and $\textbf{A}$ are statistically independent $(D\times D)$ Hermitian matrices, each sampled from GUE, with probability densities  
\begin{align}\label{AppMatrixS} 
\mathcal{P}_{S}(\textbf{S}) = 2^{-\frac{D}{2}} \left(\frac{2D}{\pi(1+\tau)v^2}\right)^{\frac{D^2}{2}} e^{-\frac{D}{(1+\tau)v^{2}}\tr\textbf{S}^2},  
\end{align}  
and  
\begin{align}\label{AppMatrixA} 
\mathcal{P}_{A}(\textbf{A}) = 2^{-\frac{D}{2}} \left(\frac{2D}{\pi(1-\tau)v^2}\right)^{\frac{D^2}{2}} e^{-\frac{D}{(1-\tau)v^{2}}\tr\textbf{A}^2},  
\end{align}  
respectively. In Eq.~\eqref{GinUEMM1}, essentially, the symmetry of the ``primary" Hamiltonian / matrix $\textbf{S}$ is broken by adding to it the Hamiltonian / matrix $i\textbf{A}$, and the degree of this symmetry breakage is regulated by the crossover parameter $\tau$. The parameter $v^{2} > 0$ controls the spectral bandwidth, setting the variance scale of the matrix elements. In particular, it determines the geometric boundary of the spectrum in the complex plane, fixing the support of eigenvalues $(z_i = x_i + i y_i)$ on the ellipse
$\frac{x^{2}}{(1+\tau)^{2}} + \frac{y^{2}}{(1-\tau)^{2}} \le v^{2}$.
For convenience, we set $v^2=1$, i.e., we measure energies in units of the spectral bandwidth, which does not affect universal properties. More generally, however, when comparing eGinUE with a physical model, $v^{2}$ can be treated as an effective bandwidth parameter. In our analysis, the Heisenberg time $T_{\mathrm{H}}(\tau,D)$ extracted from the DSFF depends on $(D / v^{2})^{\mu(\tau)}$ in a known way  [see Table~\ref{tab:TH_definitions}], where the exponent $\mu(\tau)$ encodes the $\tau$-dependent scaling. Thus, the numerically determined Heisenberg time of a physical model can be used to infer an effective value $v_{\mathrm{eff}}^{2}(\tau)$. Substituting $v_{\mathrm{eff}}^{2}(\tau)$ for $v^{2}$ in the eGinUE expressions rescales the analytic DSFF into the same spectral units as the physical system, ensuring that the mean level spacings coincide without any further fitting or manual time-axis rescaling.
With the aid of distributions of ${\bf S}$ and ${\bf A}$, one can obtain the probability density function for $\bH$ describing the eGinUE as~\cite{fyodorov1997almostcrossover},
\begin{align}{\label{GinUEMM2}}
    \mathcal{P}_{H}(\textbf{H})= \left(\frac{D}{\pi v^{2}\sqrt{1-\tau^{2}}}\right)^{D^2} e^{-\frac{D\,\tr(\textbf{H}\textbf{H}^{\dagger}-\tau \mathrm{Re}(\textbf{H}^{2}))}{(1-\tau^{2})v^2}},
\end{align}
where $\mathrm{Re}$ stands for the real part. In the limits of $\tau\to 0$ and $\tau\to 1$, we obtain GinUE and GUE, respectively, with the density function for $\bH$ reducing to $ \mathcal{P}_{H}(\textbf{H})\propto e^{-\frac{D}{v^2}\tr \bH\bH^\dag}$ and $ \mathcal{P}_{H}(\textbf{H})\propto e^{-\frac{D}{2v^2}\tr \bH^2}$, respectively.
Since computation of the DSFF involves distribution of eigenvalues, our starting point relies on utilizing the results derived in Ref.~\cite{fyodorov1997almostcrossover}. Starting from the matrix distribution in Eq.~\eqref{GinUEMM2}, the joint probability density $\mathcal{P}_{D}(\{z_{i}\}_{i=1}^D)$ of the complex eigenvalues can be derived to be~\cite{fyodorov1997almostcrossover}
\begin{align}
\mathcal{P}_{D}(\{z_{i}\}_{i=1}^D)=&\frac{D^{D(D+1)/2}}{\pi^N (1-\tau^2)^{D/2}(\prod_{m=1}^D m!)}\,\prod_{j>k}|z_j-z_k|^2\exp\left[\frac{-D}{(1-\tau^2)}\sum_{l=1}^D \big\{|z_l|^2-\frac{\tau}{2}(z_l^2+z_l^{*2}) \big\}\right].
\end{align}

The standard Dyson-Mehta procedure based on orthogonal polynomials then enables one to write down the $n$-eigenvalue correlation function $(n\leq D)$, which represents the joint probability density of finding $n$ eigenvalues, obtained by integrating out the remaining $(D-n)$ eigenvalues from the full $D$ -eigenvalue joint probability distribution
\begin{align}\label{AppenEvCorre}
 R_{n}(z_1,...,z_n)=\frac{D!}{(D-n)!}\int d^2z_{n+1}\cdots d^2z_D \mathcal{P}_{D}(\{z_{i}\}_{i=1}^D). 
\end{align}
Eq.~\eqref{AppenEvCorre} is known to have a determinent form
\begin{align}
      R_{n}(z_1,...,z_n)= \text{det}[\mathcal{K}(z_{j},z_{k}^{*})]_{j,k=1}^{n}, 
\end{align}
where the kernel $\mathcal{K}(z_j,z_k^*)$ is given by 
\begin{align}\label{kernel}
\mathcal{K}(z_j,z_k^*) = w {(z_j)} w{(z_k^{*})}\sum\limits_{n=0}^{D-1} p_n(z_j)p_n(z_k^{*}),  
\end{align}
with the weight function 
\begin{align}
w^{2}{(z)}=\exp\left[ -\frac{D}{1-\tau^{2}}\left(|z|^{2}-\frac{\tau}{2}(z^{2}+z^{*2}) \right)\right].
\end{align}
Polynomials $p_{j}(z)$ in Eq.~\eqref{kernel} are expressible in terms of scaled Hermite polynomials [$H_{j}(u)$] with complex argument~\cite{di1994laughlin,forrester1996two}, viz.,
\begin{align}\label{poly}
   p_{j}(z)=\frac{\sqrt{D}}{\sqrt{\pi\,j!}\,(1-\tau^{2})^{1/4}}\left(\frac{\tau}{2}\right)^{j/2} H_{j}\left(\sqrt{\frac{D}{2\tau}}\,z\right),
\end{align}
where $j=0,1,2,\ldots$. These polynomials in Eq.~\eqref{poly} satisfy the orthonormality relation,
\begin{align}
\int d^2z\, w^2(z) p_j(z)p_k(z^*)=\delta_{jk},
\end{align}
where $d^{2}z=|z|\,d|z|\,d(\mathrm{arg}(z))=dx\,dy$, with the integral extending over the entire complex plane.
The one-eigenvalue correlation function $R_1(z)$, by putting $n=1$ in Eq.~\eqref{AppenEvCorre}, represents the two-dimensional density of states, giving the probability density eigenvalue at position $z$ in the complex plane. By construction, integrating over the complex plane one gets the total number of eigenvalues $\int d^2 z\, R_1(z) = D.$

Starting from the definition of the DSFF, it can be computed using the two-point correlation function as
\begin{align}\label{DSFFmain}
\nonumber
&\mathcal{F}(t,s)=1+\frac{1}{D}\left\langle \sum_{n,m=1 \atop (n\ne m)}^{D}e^{it(x_{n}-x_{m})+is(y_{n}-y_{m})} \right\rangle
=1+\frac{1}{D}\int d^{2}z_{1}d^{2}z_{2}e^{it(x_{1}-x_{2})+is(y_{1}-y_{2})}R_{2}(z_{1},z_{2})\\
     \nonumber
&~~~~~~~~\equiv1+\mathcal{F}_{\text{dis}}(t,s)-\mathcal{F}_{\text{conn}}(t,s),\\
\end{align}
where,
\begin{align}\label{DSFFmain_dis}
\mathcal{F}_{\text{dis}}(t,s)=\frac{1}{D}\int d^{2}z_{1}d^{2}z_{2}e^{it(x_{1}-x_{2})+is(y_{1}-y_{2})}\mathcal{K}(z_{1},z_{1}^*)\mathcal{K}(z_{2},z_{2}^*),\\
\label{DSFFmain_conn}
    \mathcal{F}_{\text{conn}}(t,s)=\frac{1}{D}\int d^{2}z_{1}d^{2}z_{2}e^{it(x_{1}-x_{2})+is (y_{1}-y_{2})}\mathcal{K}(z_{1},z_{2}^*) \mathcal{K}(z_{2},z_{1}^*).
\end{align}
The three terms in the RHS of Eq.~\eqref{DSFFmain} correspond to the contact, disconnected ($\mathcal{F}_{\text{dis}}$), and connected ($\mathcal{F}_\text{conn}$) parts of the DSFF, respectively. The derivation of disconnected and connected parts relies on utilizing two important identities involving $\alpha^{\text{th}}$ order Hermite polynomials, viz., the addition formula (umbral identity),
\begin{align}\label{umbral}
H_{\alpha}(a+b)=\sum_{\beta=0}^{\alpha}\binom{\alpha}{\beta}H_{\beta}(a)\,(2b)^{\alpha-\beta},
\end{align}
and the corresponding orthogonality relation given by~\cite{van1990new},
\begin{align}\label{orthog}
 &\int_{-\infty}^{\infty}\int_{-\infty}^{\infty}da\, db\, H_{\alpha}(a+ib)\,H_{\gamma}(a-ib)\,\mathrm{e}^{-ua^{2}-vb^{2}}=\frac{\pi}{\sqrt{uv}}\,2^{\alpha}\,\alpha! \,\left(\frac{1}{u}+\frac{1}{v}\right)^{\alpha}\,\delta_{\alpha,\gamma},
\end{align}
where, $\mathrm{Re}(u), \mathrm{Re}(v)>0$.
In the disconnected term in Eq.~\eqref{DSFFmain}, both the integrals $z_1$ and $z_2$ lead to identical results, and we need to evaluate
\begin{align}
\mathcal{F}_{\text{dis}}=\frac{1}{D}\int d^{2}z_{1}d^{2}z_{2}e^{it(x_{1}-x_{2})+is(y_{1}-y_{2})}\mathcal{K}(z_{1},z_{1}^*)\mathcal{K}(z_{2},z_{2}^*)=\frac{1}{D}\left(\int d^{2}z\, e^{ixt+iys}\,\cK(z,z^{*})\right)^{2}.
\end{align}
Equations~\eqref{kernel}-\eqref{poly} are now employed to replace the kernel $\cK(z,z^*)$ within the above integral and give
\begin{align}
\mathcal{F}_{\text{dis}}=\frac{1}{D}\left(\sum_{n=0}^{D-1}\frac{D\tau^n}{n! \,2^n \pi (1-\tau^2)^{1/2}}I_{n,n}(\tau,t,s)\right)^2,
\end{align}
where
\begin{align}\label{hermiteint}
I_{n,m}(\tau,t,s)=&\int \,d^{2}z H_{n}\left(\sqrt{\frac{D}{2\tau}}z\right)H_{m}\left(\sqrt{\frac{D}{2\tau}}z^{*}\right)\exp\Big(ixt+iys-\frac{Dx^2}{1+\tau}-\frac{Dy^2}{1-\tau} \Big).
\end{align}
The evaluation of the integral in Eq.~\eqref{hermiteint} requires a detailed algebraic calculation aided by Eqs.~\eqref{umbral},~\eqref{orthog}, and yields,
\begin{align}\label{FdisSum}
  \mathcal{F}_{\text{dis}}(t,s)&= \frac{e^{-\frac{(1+\tau)t^{2}}{2\,D}-\frac{(1-\tau)s^{2}}{2\,D}}}{D}\left[\sum_{n=0}^{D-1}\sum_{k=0}^{n}\binom{n}{k}\frac{(-\mathbb{T})^{n-k}}{(n-k)!}\right]^{2},
\end{align}
where $\mathbb{T}$ is defined as
\begin{align}
\label{scrT}
\mathbb{T}=\frac{(1+\tau)^{2}t^{2}+(1-\tau)^{2}s^{2} }{4\,D}=\frac{\eta^2}{D}|T|^2, \quad \textrm{with} \quad \eta=\frac{1}{2}[(1+\tau)^{2}\cos^2\varphi+(1-\tau)^{2}\sin^2\varphi]^{1/2}, 
\end{align}
where $\varphi$ is the argument of $T$. We note that for $\tau=0$ and $\tau=1$, $\eta$ reduces to $1/2$ and $\cos\varphi$, respectively. The inner sum (over $k$) in Eq.~\eqref{FdisSum} is identified as the expansion of Laguerre polynomial $L_n(z)\equiv L_n^{(0)}(z)$. Hence we have,
\begin{align}
\mathcal{F}_{\text{dis}}(t,s)= \frac{1}{D}e^{-\frac{(1+\tau)t^{2}}{2\,D}-\frac{(1-\tau)s^{2}}{2\,D}}\left[\sum_{n=0}^{D-1} L_n^{(0)}(\mathbb{T})\right]^{2}.
  \end{align}
Finally, the sum over Laguerre polynomials can be performed using the relation $\sum_{a=0}^b L_a^{(k)}(z)=L_b^{(k+1)}(z)$, which leads to 
\begin{align}{\label{Fdis}}
  \mathcal{F}_{\text{dis}}(t,s)=\frac{1}{D}e^{-\frac{(1+\tau)t^{2}}{2\,D}-\frac{(1-\tau)s^{2}}{2\,D}}\big[L_{D-1}^{(1)}(\mathbb{T})\big]^{2}.
\end{align}
In Eq.~\eqref{Fdis}, instead of associated Laguerre polynomial, we may use the Kummer confluent hypergeometric function $M(a,b,z)=\!\,_1F_1(a,b,z)$, or the Tricomi confluent hypergeometric function $U(a,b,z)$ using the relation, 
\begin{align}
    L_a^{(k)}(z)=\binom{a+k}{a}M(-a,k+1,z)=\frac{(-1)^a}{a!}U(-a,k+1,z),
\end{align}
which holds for non-negative integer $a$, with $\binom{u}{v}$ representing the binomial coefficient. 
 
Next, we aim to evaluate the connected part $\cF_\mathrm{conn}(t,s)$, following similar initial steps as in the derivation of $\cF_\mathrm{dis}(t,s)$, and find that
\begin{align}{\label{A4}}
\cF_{\mathrm{conn}}(t,s)=\frac{1}{D}\int d^{2}z_{1} d^{2}z_{2}\, e^{i\,(x_{1}-x_{2})t+i(y_{1}-y_{2})s}\,|\cK(z_{1},z_{2}^{*})|^{2}
= \frac{D}{\pi^{2}(1-\tau^2)}\sum_{n,m=0}^{D-1}\frac{(\tau/2)^{n+m}}{n!\,m! }\,I_{n,m}(\tau,t,s)\,I_{n,m}^{*}(\tau,t,s).
\end{align}
Again, a lengthy algebra results in
\begin{equation}
\mathcal{F}_{\text{conn}}(t,s)=\frac{1}{D}e^{-\frac{(1+\tau)t^{2}}{2\,D}-\frac{(1-\tau)s^{2}}{2\,D}}\sum_{n,m=0}^{D-1}\frac{(4\,D)^{-(n+m)}}{n!\,m!}
\times \left|\sum_{k=0}^{\min(n,m)}\binom{n}{k}\binom{m}{k}\frac{k!}{(4\,D)^{-k}} \frac{\left[i(1+\tau)t-(1-\tau)s\right]^{n-k}}{\left[i(1+\tau)t+(1-\tau)s\right]^{k-m}} \right|^{2}.
\end{equation}
Mathematica~\cite{Mathematica} is able to perform the inner double-sum in terms of Tricomi confluent hypergeometric function, which can be equivalently written in terms of associated Laguerre polynomial, as 
\begin{align}{\label{Fconn}}
\cF_{\text{conn}}(t,s)=&\frac{1}{D}e^{-\frac{(1+\tau)t^{2}}{2\,D}-\frac{(1-\tau)s^{2}}{2\,D}}\sum_{n,m=0}^{D-1}\frac{m!}{n!}\mathbb{T}^{n-m}\big[L_m^{(n-m)}(\mathbb{T})\big]^{2}.
\end{align}
The double sum over $n,m$ in Eq.~\eqref{Fconn}  may be separated using $n=m$ and $n>m$ (or $n<m$) parts only by using the following property of the associated Laguerre polynomials: $L_a^{(-k)}(z)=(-z)^k \frac{(a-k)!}{a!}L_{a-k}^{(k)}(z)$.

Therefore, we now have 
\begin{align}\label{DSFF_exact}
    \mathcal{F}(t,s)=1+\frac{1}{D}e^{-\frac{(1+\tau)t^{2}}{2\,D}-\frac{(1-\tau)s^{2}}{2\,D}}\big[L_{D-1}^{(1)}(\mathbb{T})\big]^{2}-\frac{1}{D}e^{-\frac{(1+\tau)t^{2}}{2\,D}-\frac{(1-\tau)s^{2}}{2\,D}}\sum_{n,m=0}^{D-1}\frac{m!}{n!}\mathbb{T}^{n-m}\big[L_m^{(n-m)}(\mathbb{T})\big]^{2}.
\end{align}
This expression represents the exact analytical form of the DSFF for the eGinUE.

To verify the consistency of our exact result, it is instructive to check that it reproduces the known expressions for the limiting cases $\tau=0$ and $\tau=1$, corresponding to the GinUE and GUE, respectively. Starting from Eq.~\eqref{DSFF_exact}, we first consider the fully non-Hermitian limit, $\tau=0$. In this case, the expression simplifies to the well-established GinUE form~\cite{PhysRevLett.127.170602,garcia2023universality}
\begin{align}
\label{Fginue_supp}
\mathcal{F}^\mathrm{GinUE}(|T|=t^{2}+s^{2})= 1+\frac{e^{-\frac{|T|^{2}}{2\,D}}}{D}\left[ L_{D-1}^{(1)}\bigg(\frac{|T|^2}{4D}\bigg)\right]^{2}-\frac{e^{\frac{|T|^{2}}{2\,D}}}{D}\sum_{n,m=0}^{D-1}\frac{m!}{n!}\bigg(\frac{|T|^{2}}{4D}\bigg)^{n-m}\left[L_m^{(n-m)}\bigg(\frac{|T|^{2}}{4\,D}\bigg)\right]^{2}.
\end{align}
In the opposite limit, $\tau=1$, corresponding to the Hermitian case, we recover the exact GUE expression~\cite{del2017scrambling} for the SFF
\begin{align}\label{Fgue_supp}
\mathcal{F}^\mathrm{GUE}(t)= 1+\frac{e^{-\frac{t^{2}}{D}}}{D}\left[L_{D-1}^{(1)}  \bigg(\frac{t^2}{D}\bigg)   \right]^{2}-\frac{e^{-\frac{t^{2}}{D}}}{D}\sum_{n,m=0}^{D-1}\frac{m!}{n!}\left(\frac{t^{2}}{D}\right)^{n-m}\left[L_m^{(n-m)}\bigg(\frac{t^{2}}{D}\bigg)\right]^{2}.
\end{align}
These limiting cases not only serve as non-trivial checks of our general expression but also demonstrate that the eGinUE interpolates smoothly between GinUE ($\tau=0$) and GUE ($\tau=1$) through the non-Hermiticity parameter $\tau$.

A close inspection of the limiting expressions [Eqs.~\eqref{Fginue_supp} and~\eqref{Fgue_supp}] reveals a remarkable ``scaling relation'' connecting the GinUE and GUE results through their disconnected (second term) and connected (third term) parts, given by
\begin{align}
\label{FGinGUscale_supp}
\mathcal{F}^\mathrm{GinUE}_\mathrm{dis,conn}(t,s)=e^{\frac{-|T|^2}{4D}}\mathcal{F}^\mathrm{GUE}_\mathrm{dis,conn}(|T|/2).
\end{align}
Analogously, for eGinUE, the DSFF [Eq.~\eqref{DSFF_exact}] can be expressed in terms of the GUE result [Eq.~\eqref{Fgue_supp}] as
\begin{align}
\label{scaling_supp}
\cF_\mathrm{dis}(t,s)=e^{-\frac{(1-\tau^2)(t^2+s^2)}{4D}} \mathcal{F}^\mathrm{GUE}_\mathrm{dis}(\sqrt{D\mathbb{T}}) \quad \textrm{and} \quad \cF_\mathrm{conn}(t,s)=e^{-\frac{(1-\tau^2)(t^2+s^2)}{4D}} \mathcal{F}^\mathrm{GUE}_\mathrm{conn}(\sqrt{D\mathbb{T}})
\end{align}
The scaling relation in Eq.~\eqref{scaling_supp} is immensely powerful. By leveraging known results from GUE, one can infer corresponding results for eGinUE, which proves particularly useful in deriving large-$D$ approximate results for the DSFF.

\section{Derivation of large-$D$ DSFF results}\label{AppDSFFAsy}

Starting from the exact analytical expressions derived in Sec.~\ref{AppDSFF}, we now obtain the asymptotic form of the DSFF of eGinUE by employing suitable approximations valid in the large-$D$ limit. For the disconnected part [Eq.~\eqref{Fdis}], we use the following asymptotic form of the associated Laguerre polynomials~\cite{szeg1939orthogonal}, which holds for large-$D$ 
\begin{align}
L_D^{(\alpha)}(x/D) = D^{\alpha} x^{-\alpha/2} J_{\alpha}(2x^{1/2}),
\end{align}
with fixed $\alpha$ and $x>0$. Here, $J_{\alpha}(u)$ denotes the Bessel function of order $\alpha$. This readily leads to the large-$D$ approximation of the disconnected part [Eq.~\eqref{Fdis}] as
\begin{align}
\label{FdisAsy}
\mathcal{F}_\mathrm{dis}(t,s)&\approx\frac{1}{\mathbb{T}}\,e^{-\frac{(1-\tau^2)(t^2+s^2)}{4D}}J_1\big(2\sqrt{D\mathbb{T}}\big)^{2}.
\end{align}

The connected part of the DSFF is more involved due to the double summation appearing in Eq.~\eqref{Fconn}. However, for the connected part, a very good approximation for the SFF (equivalently DSFF) of GUE has been derived in Ref.~\cite{liu2018spectral}, given by 
\begin{align}\label{FuncR}
\mathcal{F}^{\mathrm{GUE}}_\mathrm{conn}(t)\approx R(t)\quad \textrm{where} \quad
    R(t)=
\begin{cases} 
\frac{2}{\pi}\csc^{-1}\left(\frac{2D}{\sqrt{4D^2-t^2}}\right)-\frac{t \sqrt{4D^2-t^2}}{2\pi D^2}, & t< 2D \\
0, & t\ge 2D.
\end{cases}
\end{align}
Based on the function~\eqref{FuncR} and the scaling relation connecting the eGinUE DSFF to the GUE SFF result~\eqref{scaling_supp}, one can at once write down the large-$D$ approximation for the connected part [Eq.~\eqref{Fconn}] of the eGinUE DSFF, as 
\begin{align}
\label{FconnAsy}
\mathcal{F}_\mathrm{conn}(t,s)\approx e^{-\frac{(1-\tau^2)(t^2+s^2)}{4D}}R\big(\sqrt{D\mathbb{T}}\big).
\end{align}
The final large-$D$ approximation of the eGinUE DSFF, obtained by combining the individual term-wise results, is
\begin{align}{\label{FAsy}}
   \mathcal{F}(t,s)\approx 1 + \frac{1}{\mathbb{T}}\,e^{-\frac{(1-\tau^2)(t^2+s^2)}{4D}}J_1\big(2\sqrt{D\mathbb{T}}\big)^2-e^{-\frac{(1-\tau^2)(t^2+s^2)}{4D}}R\big(\sqrt{D\mathbb{T}}\big).
\end{align}
In the $\tau=0$ and $\tau=1$ limits, we readily obtain the large-$D$ approximations for the DSFF as follows: for GinUE~\cite{PhysRevLett.127.170602}
\begin{align}\label{FGinUEAsy}
\mathcal{F}^{\mathrm{GinUE}}(t,s) \approx 1 + e^{-\frac{|T|^2}{4D}} \frac{4D}{|T|^2} J_1(|T|)^2
- e^{-\frac{|T|^2}{4D}} R(|T|/2) \approx 1 + \frac{4D}{|T|^2} J_1(|T|)^2 - e^{-\frac{|T|^2}{4D}},
\end{align}
and for GUE~\cite{liu2018spectral},
\begin{align}\label{FGUEAsy}
\mathcal{F}^{\mathrm{GUE}}(t) &\approx 1 + \frac{D}{t^2} J_1(2t)^2 - R(t).
\end{align}

\begin{figure}
\includegraphics[width=0.6\columnwidth]{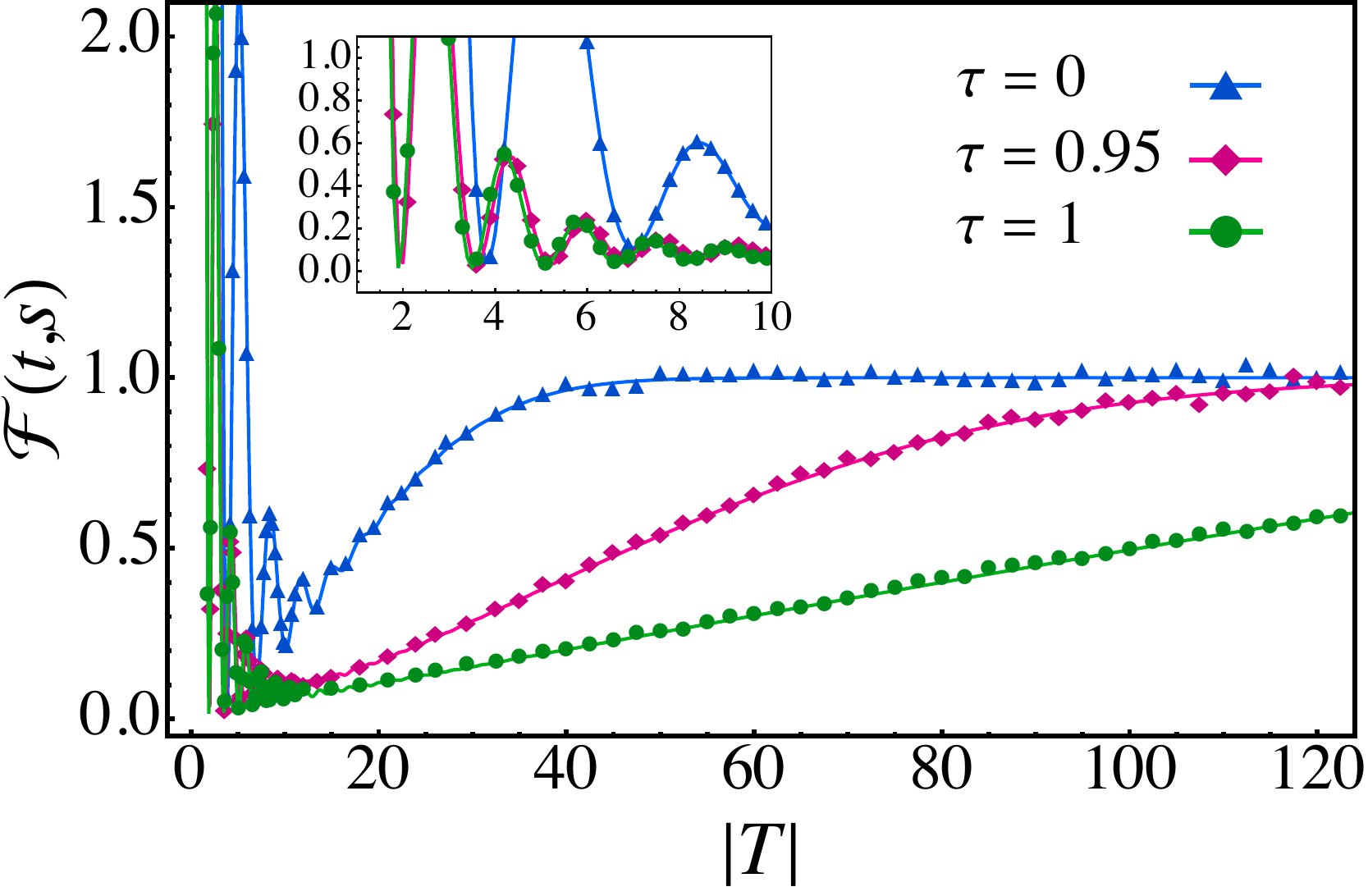}
\caption{Plots of the eGinUE DSFF $\mathcal{F}(t,s)$ as a function of $|T|$ for $D=125$ and $\varphi=0$. The solid curves represent large-$D$ approximate analytical results [Eq.~\eqref{FAsy}], while symbols are based on Monte Carlo simulations of the eGinUE matrix model. The inset provides an enlarged view of the \emph{dip} region, demonstrating that the large-$D$ approximation accurately captures even the oscillations. }
\label{fig2}
\end{figure}

In Fig.~\ref{fig2}, we compare the large-$D$ analytical approximation [Eq.~\eqref{FAsy}] with the Monte Carlo simulation of the eGinUE matrix model for $D=125$ and $\varphi=0$, averaged over 5000 realizations. The agreement is excellent — the analytical results not only reproduce the overall behavior but also accurately capture the fine oscillations observed in the \emph{dip} region.

\section{Refined estimates of $T_\mathrm{H}$ and $T_\mathrm{Th}$}
\label{AppTime}
In the main text, we present order estimates and refined analytical expressions for $T_\mathrm{H}$ and $T_\mathrm{Th}$. Here, we revisit these results and present their derivation in detail, including the underlying approximations and assumptions.

\subsection{Heisenberg time $T_\mathrm{H}$}\label{subsec_TH}
To find the refined estimates of $T_\mathrm{H}$, we consider the condition $\mathcal{F}_\mathrm{conn} [\textrm{Eq.}~\eqref{DSFFmain_conn}] \approx 0$. In the strong non-Hermiticity regime, $\tau < \tau_c$, with $\tau_c = 1 - c/D$ (as defined in the main text, where $c \sim \mathcal{O}(1)$), the exponential term in Eq.~\eqref{FconnAsy} causes $\mathcal{F}_\mathrm{conn}$ to decay much more rapidly than the factor $R(\sqrt{D\mathbb{T}})$, which remains $\mathcal{O}(1)$. To obtain a reasonable estimate for $T_\mathrm{H}$ in this regime, we consider the asymptotic decay form of the connected part
\begin{align}\label{Fconn_decay}
\mathcal{F}_\mathrm{conn}(t,s) &\approx \exp\left(-\frac{(1 - \tau^2)(t^2 + s^2)}{4D} \right) = \exp\left(-\frac{(1 - \tau^2)|T|^2}{4D} \right).
\end{align}
If one were to set $\mathcal{F}_\mathrm{conn} = 0$ directly, this would imply $T_\mathrm{H} \to \infty$, which is neither practical nor informative in this regime. To avoid this divergence, instead we define $T_\mathrm{H}$ as the time at which $\mathcal{F}_\mathrm{conn}$ reaches a small finite value $\epsilon$, chosen to characterize the decay scale in this limit. As a convenient choice, we take $\epsilon = e^{-4} \approx 0.018$, resulting in $T_\mathrm{H} \approx 4\sqrt{D/(1 - \tau^2)}$. Although the choice of $\epsilon$ is heuristic, the resulting estimate consistently aligns well with the observed onset of the plateau for various values of $\tau$ within this regime (see Fig.~\ref{fig3} in the main text). An alternative yet complementary approach is to define $T_\mathrm{H}$ in terms of the mean level spacing $\delta$ between adjacent eigenvalues, $T_\mathrm{H} = 2\pi/\delta$~\cite{haake2010quantum}. This definition is universally applicable to quantum systems with discrete spectra—integrable, chaotic, or periodically driven—and corresponds to the time scale at which spectral discreteness becomes resolvable. In two-dimensional spectra, the mean level spacing scales as $\delta \sim \rho^{-1/2}$, where $\rho$ denotes the spectral density in the bulk. For eGinUE, in the large-$D$ limit, the bulk density in the strong non-Hermiticity regime is given by $\rho = D/\big(\pi(1 -\tau^2)\big)$~\cite{fyodorov1998universality}, which corresponds to the number of eigenvalues per unit area within the elliptic support of the spectrum. This yields $T_\mathrm{H} \approx 2\sqrt{\pi}\sqrt{D/(1 - \tau^2)}\approx3.54\sqrt{D/(1 - \tau^2)}$. Unlike the previous approach, this estimate does not rely on heuristic assumptions and provides an analytically clean expression for $T_\mathrm{H}$ in the $\tau < \tau_c$ regime. Both of these approaches lead to scaling $T_\mathrm{H}\sim\mathcal{O}(D^{1/2})$ in the GinUE case.

In contrast, in the weak non-Hermiticity regime around and beyond $\tau = \tau_c$, the condition $\mathcal{F}_\mathrm{conn} \approx 0$ serves as an effective approximation for estimating $T_\mathrm{H}$. Expanding $\cF_\mathrm{conn}$ [Eq.~\eqref{FconnAsy}] up to $\mathcal{O}(1/D)$ gives
\begin{align}\label{Fconn_expand}
\mathcal{F}_\mathrm{conn}(t,s)\approx1-\frac{2\eta|T|}{\pi\,D}-(1-\tau^{2})\frac{|T|^{2}}{4\,D}.
\end{align}
Setting Eq.~\eqref{Fconn_expand} to zero yields $T_{\mathrm{H}}\approx \pi D/(2\eta)$ for $\tau \approx \tau_c$. 
In the regime $\tau_c < \tau \leq 1$, $\cF_\mathrm{conn}$ [Eq.~\eqref{FconnAsy}] vanishes identically due to the factor $R(\sqrt{D\mathbb{T}})$. In this regime, the Heisenberg time $T_\mathrm{H} \approx 2D/\eta$ follows directly from the definition of $R(t)$ [Eq.~\eqref{FuncR}] and remains consistent with the order estimate $T_\mathrm{H} \sim \mathcal{O}(D)$ known for the GUE case. Alternatively, $T_\mathrm{H}$ in these regimes can also be estimated from the relation $T_\mathrm{H} = 2\pi/\delta$, using the expressions available for $\rho$~\cite{fyodorov1998universality,fyodorov1997almosteigenvaluedensity}. However, in contrast to the $\tau<\tau_c$ regime, these expressions are typically integral representations and less explicit, hence less precise for refined estimates. Note that Eq.~\eqref{Fconn_expand} captures the local behavior of the \emph{ramp}. In particular, this ramp exhibits quadratic growth in the GinUE limit and linear growth in the GUE limit.

\subsection{Thouless time $T_\mathrm{TH}$}\label{subsec_TTH}
\begin{figure}
\includegraphics[width=0.5\columnwidth]{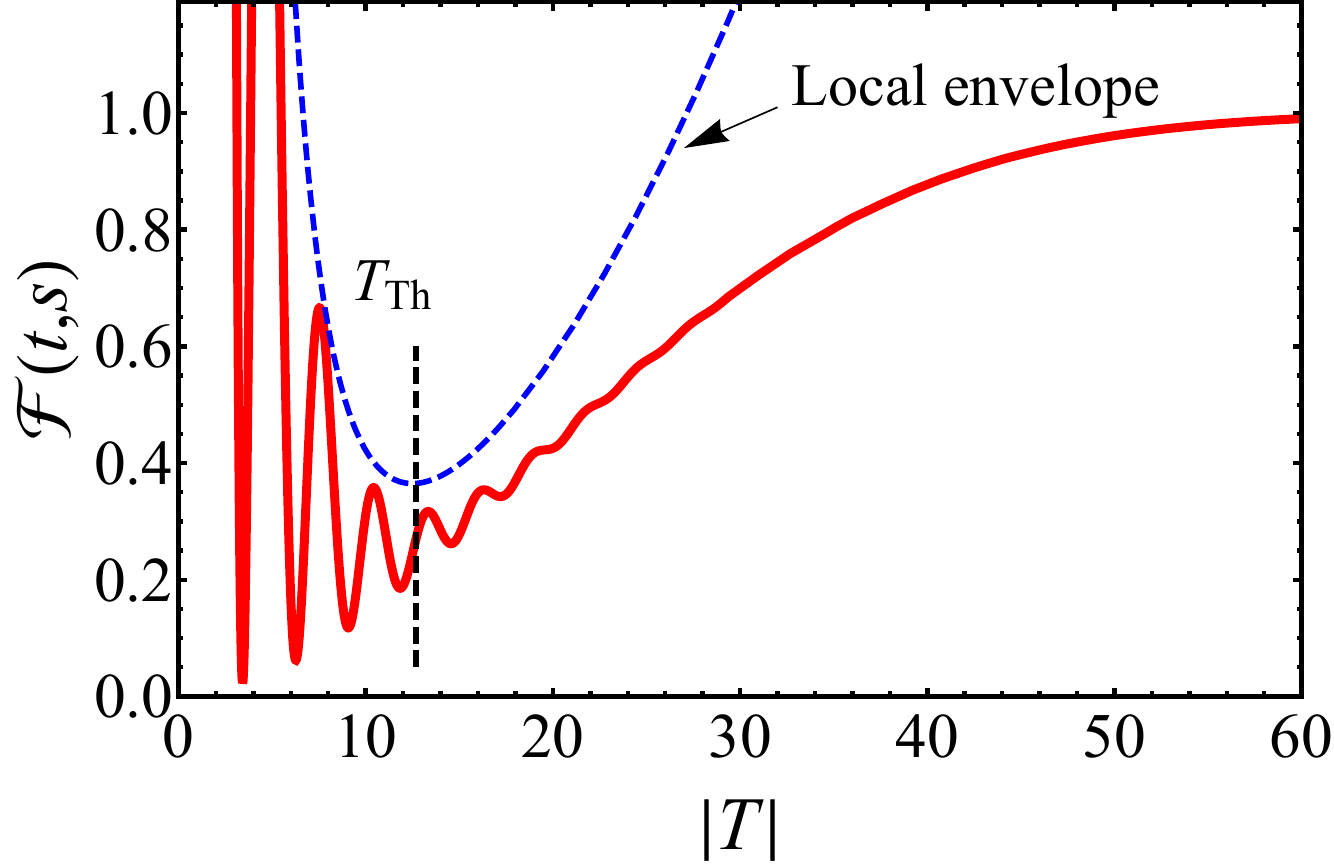}
\caption{Plot of the DSFF curve (solid red line) of eGinuE using Eq.~\eqref{FAsy}. The local envelope, Eq.~\eqref{Envelope} (dashed blue), is overlaid for comparison. The Thouless time $T_\mathrm{Th}$ (vertical black dashed line) is determined by solving Eq.~\eqref{Quintic} for $|T|$. We choose here $D = 150$, $\varphi = \pi/4$, and $\tau = 0.5$.
}
\label{figAppen2}
\end{figure}

We now look at the behaviour of the Thouless time, $T_{\mathrm{Th}}$. For time values in the neighborhood of $ T_\mathrm{Th} $, to extract the leading $|T|$-dependence of the disconnected term $ \mathcal{F}_\mathrm{dis}$ [Eq.~\eqref{FdisAsy}], we consider time scales of the form $ |T| \sim \mathcal{O}(D^\alpha) $, where $ 0 < \alpha < 1 $, and then expand in $ |T| $ for large-$D$. The lower bound $ \alpha > 0 $ ensures that we move beyond the short-time regime dominated by non-universal features specific to model details.
Given that the Heisenberg time scales as $ T_\mathrm{H} \sim D^{\beta} $ with $ \beta \leq 1 $ for the eGinUE, the upper bound $ \alpha < \beta $ constrains the time scale to remain below $ T_\mathrm{H} $.
The large-$D$ expansion of disconnected term [Eq.~\eqref{FdisAsy}] leads to
\begin{align}\label{Fdis_asymp}
\mathcal{F}_\mathrm{dis}(t,s) \approx 
\frac{D}{\pi \eta^3 |T|^3}
e^{-\frac{(1-\tau^2)|T|^2}{4D}}
\cos^2\!\left(2\eta |T|+\frac{\pi}{4}\right)
\approx 
\frac{D}{\pi \eta^3 |T|^3}
\cos^2\!\left(2\eta |T|+\frac{\pi}{4}\right).
\end{align}
To estimate $T_\mathrm{Th}$, we consider the local envelope [$\mathcal{F^{\mathrm{Enve}}}(t,s)$] of the DSFF curve, constructed from the combination of the leading behavior of the connected term [Eq.~\eqref{Fconn_expand}] and the
non-oscillatory factor \big($\frac{D}{\pi \eta^3 |T|^3}$\big) of the disconnected term [Eq.~\eqref{Fdis_asymp}] 
\begin{align}\label{Envelope}
\mathcal{F^{\mathrm{Enve}}}(t,s) = 1 + \frac{D}{\pi \eta^3 |T|^3} - \bigg(1-\frac{2\eta |T|}{\pi D}-\frac{(1-\tau^2)|T|^2}{4D}\bigg)=\frac{(1 - \tau^2)|T|^2}{4D} + \frac{2\eta\,|T|}{\pi D} + \frac{D}{\pi \eta^3 |T|^3}.
\end{align}
The Thouless time $T_\mathrm{Th}$ is then estimated by locating the minimum of $\mathcal{F^{\mathrm{Enve}}}(t,s)$ (see Fig.~\ref{figAppen2}), resulting in a quintic equation
\begin{align}\label{Quintic}
(1-\tau^2)|T|^5+\frac{4\eta}{\pi}|T|^4-\frac{6D^2}{\pi \eta^3}=0.
\end{align}
Eq.~\eqref{Quintic} predicts $T_\mathrm{Th}\sim\mathcal{O}(D^{2/5})$ and $\mathcal{O}(D^{1/2})$ for GinUE and GUE, respectively, in agreement with earlier works~\cite{PhysRevLett.127.170602,del2017scrambling}. In fact, for $\tau<\tau_c$, $T_\mathrm{Th}\approx(\frac{6}{\pi \eta^3 (1-\tau^2)})^{1/5} D^{2/5}$ serves as an excellent estimate. On the other hand, for $\tau\gtrsim \tau_c$, we have $T_\mathrm{Th}\approx (\frac{3}{2})^{1/4}\frac{D^{1/2}}{\eta}$.

In Table~\ref{tab:TH_definitions} we summarize the findings of this section for the refined estimates of $T_\mathrm{H}$ and $T_\mathrm{Th}$ in different regimes of $\tau$.
\begin{table}[h]
    \footnotesize 
    \centering
    \begin{tabular}{|c|c|c|}
        \hline
        $\tau$ regime & $T_\mathrm{H}$ & $T_\mathrm{Th}$ \\
        \hline
        $\tau < \tau_c$ & 
        $T_{\mathrm{H},1} \approx 4\sqrt{D/(1 - \tau^2)}$ (obtained from $\mathcal{F}_\mathrm{conn} \approx \epsilon = e^{-4}$) & \\
        & $\widetilde{T}_{\mathrm{H},1}
 \approx 2\pi\sqrt{D/(\pi(1 - \tau^2))}$ (obtained from $T_\mathrm{H} = 2\pi/\delta$) 
        & $T_{\mathrm{Th},1} \approx \left( \frac{6}{\pi \eta^3 (1 - \tau^2)} \right)^{1/5} D^{2/5}$ \\
        \hline
        $\tau \approx \tau_c$ & 
        $T_{\mathrm{H},2} \approx \pi D/(2\eta)$ (obtained from $\mathcal{F}_\mathrm{conn} \approx 0$) 
        & \multirow{2}{*}{$T_{\mathrm{Th} ,2}\approx \left( \frac{3}{2} \right)^{1/4} \frac{D^{1/2}}{\eta}$} \\
        \cline{1-2}
        $\tau > \tau_c$ & 
        $T_{\mathrm{H},3} \approx 2D/\eta$ (obtained from $\mathcal{F}_\mathrm{conn} \approx 0$) & \\
        \hline
    \end{tabular}
    \caption{Refined estimates of Heisenberg time $T_\mathrm{H}$ and Thouless time $T_\mathrm{Th}$ across different regimes of the non-Hermiticity parameter $\tau$.}
    \label{tab:TH_definitions}
\end{table}


\section{Additional Details on the Application of Analytical Results}\label{Application}

In this section, we present additional discussions that demonstrate the application of our analytical findings to the crossover Sachdev-Ye-Kiteav (cSYK) and crossover Power-Law Banded random matrix model (cPLBRM), complementing the analysis discussed in the main text. Before discussing the specific results for these two models, it is useful to recall a general aspect of crossover ensembles that will be important for the analysis that follows. In the thermodynamic limit, the intermediate statistics of a crossover ensemble disappear, and the system converges to one of the extreme universality classes (GinUE or GUE in this case), exhibiting a sharp transition from one class to the other. For large but finite systems, finite-size effects allow one to distinguish different symmetry classes through a continuous crossover regime, rather than letting the spectrum collapse immediately into one of the two limiting cases. Thus, the crossover can be resolved through finite-size statistics by tuning the symmetry parameter, a mechanism applicable to both cSYK and cPLBRM models. This behaviour also follows from the scaling of $\tau_c$, since $\tau_c = 1 - 1/D$ remains slightly below $1$ for any finite $D$. Only in the limit $D \to \infty$ does $\tau_c \to 1$, leaving us with two asymptotic cases $\tau = 0$ (GinUE) and $\tau = 1$ (GUE).

\subsection{DSFF of the cSYK Model in the Large-Dimensional Limit}
In the main text, it was shown that the large but finite-dimensional cSYK model provides an ideal platform to test the asymptotic analytical results derived for the eGinUE ensemble. Here, we extend this analysis tothe large-dimensional limit of the cSYK model to further  assess the validity of these results and to explore how increasing system dimension influences the crossover and universality of the DSFF. 

\begin{figure}
\includegraphics[width=0.9\columnwidth]{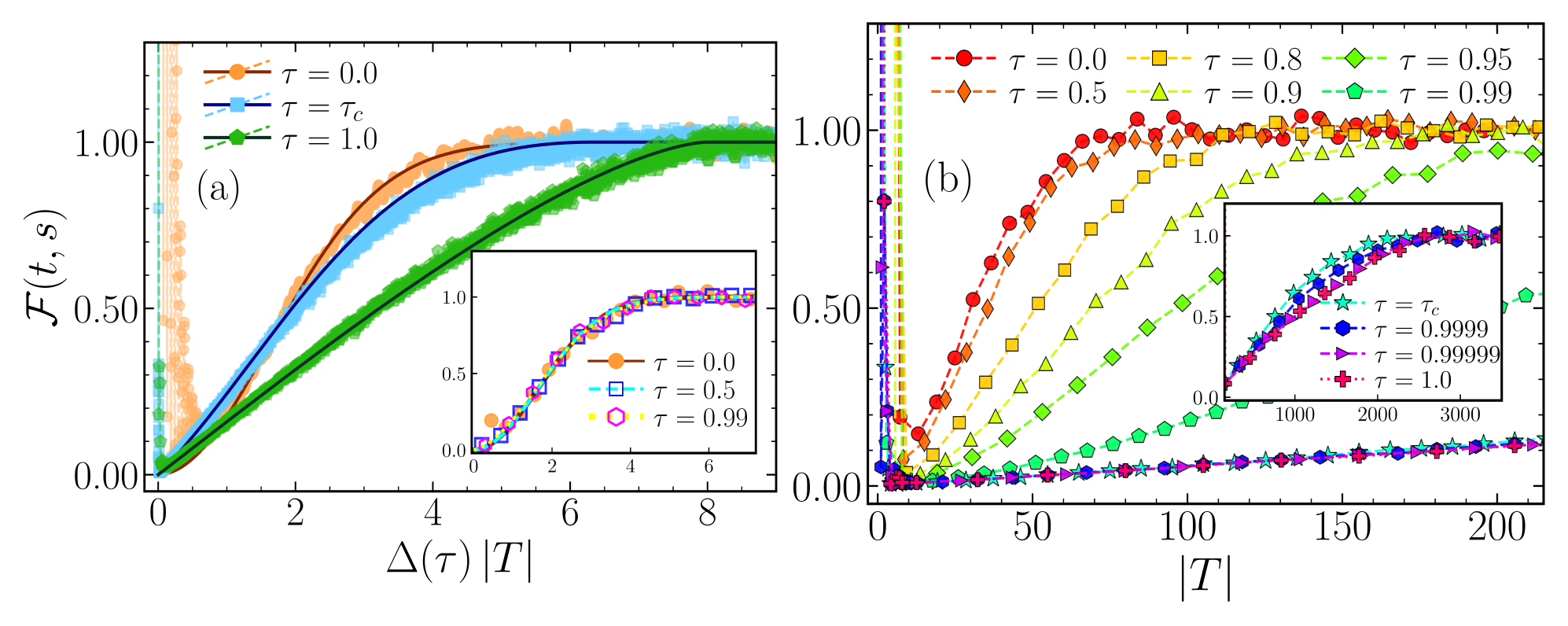}
\caption{We show the DSFF results for the $q = 4$ cSYK model with $N = 26$, normalized by the number of non-degenerate eigenvalues $2^{N/2-1}$ and averaged over $3\times10^{3}$ matrices to reduce statistical fluctuations. 
(a) Comparison with the connected part of the large-$D$ eGinUE asymptotic result (solid line), $1 - \mathcal{F}_{\mathrm{conn}}$ [Eq.~\eqref{FconnAsy}], using $D=2^{12}$ at $\varphi = 0$. 
The crossover from the GinUE to the GUE regime is evident as $\tau$ varies, with $\tau_c = 1 - 1/2^{12}$. For $\tau = 1$, we removed the edge eigenvalues and those lying very close to the symmetry axes from the cSYK spectrum, and compared the resulting data with the eGinUE analytical results containing the same number of eigenvalues.
The inset shows the same comparison for $\tau < \tau_c$, with solid, dashed, and dotted curves representing the eGinUE analytics. 
(b) DSFF data of the cSYK model over a broad range of $\tau$. The inset highlights the behavior for $\tau \gtrsim \tau_c$ at longer times.
}
\label{figAppen3}
\end{figure}

Figure~\ref{figAppen3}(a) illustrates the comparison between the DSFF of the cSYK model ($q=4$, $N=26$) and the asymptotic analytical result, $1 - \mathcal{F}_\mathrm{conn}$ [Eq.~\eqref{FconnAsy}], obtained for eGinUE matrix model ($D=2^{12}$). Both datasets are shown as functions of the rescaled time variable $\Delta(\tau)\,|T|$, where $\Delta(\tau)$ is the mean level spacing obtained numerically from the respective spectra. This rescaling expresses the time in units of Heisenberg time $T_\mathrm{H} \sim \mathcal{O} (1/\Delta(\tau))$, allowing a fair comparison across different values of $\tau$. In other words, the scaled time axis provides a common reference scale even when the spectral densities vary with $\tau$. The inset shows the collapse of all curves onto the GinUE result for $\tau$ values in the strong non-Hermitian regime ($\tau < \tau_c$), a behavior that arises due to the large system size. This aligns with the earlier discussion that larger system sizes suppress crossover features by driving the ensemble toward a universal limit.

The large dimension of the Hilbert space pushes $\tau_c$ very close to 1. Beyond $\tau_c$, the DSFF curves become highly sensitive to $\tau$ and gradually approach the Hermitian limit ($\tau = 1$). To illustrate this behavior, Fig.~\ref{figAppen3}(b) presents the raw DSFF data for several $\tau$ values. The curves for $\tau_c \lesssim \tau \leq 1$ nearly overlap at early times, while the inset reveals a small but discernible separation at longer times. In contrast, when the time axis is rescaled by the mean level spacing $\Delta(\tau)$ as in Fig.~\ref{figAppen3}(a), i.e., when time is expressed in units of Heisenberg time $T_H = 2\pi /\Delta(\tau)$, these subtle differences become more pronounced. This scaling illustrates that for larger system sizes, the DSFF becomes extremely sensitive to even small changes in $\tau_c \lesssim \tau \leq 1$.

The behavior of the DSFF at long times is further affected by the spectral properties of the cSYK system, particularly for large $N$. For $\tau_c < \tau \le 1$, a slight deviation is observed between the numerical simulations of cSYK and the analytical results of eGinUE around the Heisenberg time. To the best of our knowledge, no numerical inaccuracies or insufficient ensemble averaging accounts for this discrepancy. For large $N$, the accumulation of edge eigenvalues and the presence of near-zero eigenvalues in the cSYK spectrum significantly affect the squared summation over the eigenvalue phases underlying DSFF/SFF, producing these deviations (however, in Ref.~\cite{PhysRevD.103.106002}, this kind of behavior is addressed as finite-size effect). We find that. removing these eigenvalues effectively eliminates the discrepancy. Fig.~\ref{figAppen3}(a) shows the DSFF for $\tau = 1$ after this removal, compared to the analytical results of eGinUE using the same number of eigenvalues. However, at finite $N$, the bulk of the spectrum dominates the squared summation, making edge contributions negligible and near-zero eigenvalues insignificant. Hence, no such deviation is observed (see the $q=4, N=22$ cSYK model in Fig.~\ref{figsyk} in the main text) for finite $N$ even with relatively few realizations.
%

\subsection{Analysis of the DSFF behaviour in the cPLBRM model}
To understand the crossover behaviour of the DSFF in the cPLBRM model and its connection to the analytical results, we analyse how the parameter $\tau$ controls the degree of non-Hermiticity and influences the resulting spectral characteristics. Recall that the cPLBRM model is defined by  
\begin{align}\label{cplbrm_supp}
\textbf{H}_{ij}^{\mathrm{cPLBRM}} = \epsilon_i\,\delta_{ij} + g_{ij},
\end{align}
where $\epsilon_{i}$ and $g_{ij}=g_{ji}^*$ are complex random variables selected from the zero-mean uniform distributions with amplitudes
\begin{align}
\left|\mathrm{Re}(\epsilon_{i})\right|\leq\frac{(1+\tau)}{2}W,\,
\left|\mathrm{Im}(\epsilon_{i})\right|\leq\frac{(1-\tau)}{2}W,
\label{cPLBRM_epsilon_supp}\\
\left|\mathrm{Re}(g_{ij})\right|,\left|\mathrm{Im}(g_{ij})\right|\leq [2(|i-j|^{2}+b^{2})]^{-p/2}.
\label{cPLBRM_g_supp}
\end{align}
The bounds on onsite energies $\epsilon_{i}$ are dictated by the disorder strength $W$, while the off-diagonal couplings decay as a power law with exponent $p$ and the bandwidth of decay is controlled by $b$. 
The parameter $\tau$ controls the asymmetry between the real and imaginary parts of the diagonal elements $\epsilon_i$, thereby tuning the degree of non-Hermiticity: $\tau = 1$ suppresses the imaginary component and yields a Hermitian ensemble, whereas $\tau < 1$ introduces complex onsite terms that make the ensemble non-Hermitian. Note that the diagonal elements are complex solely due to $\epsilon_i$; the contribution from $g_{ii}$ is real in this construction. The off-diagonal couplings, on the other hand, remain Hermitian and decay with distance according to a fixed bandwidth and power-law profile. 

\begin{figure}
\includegraphics[width=0.9\columnwidth]{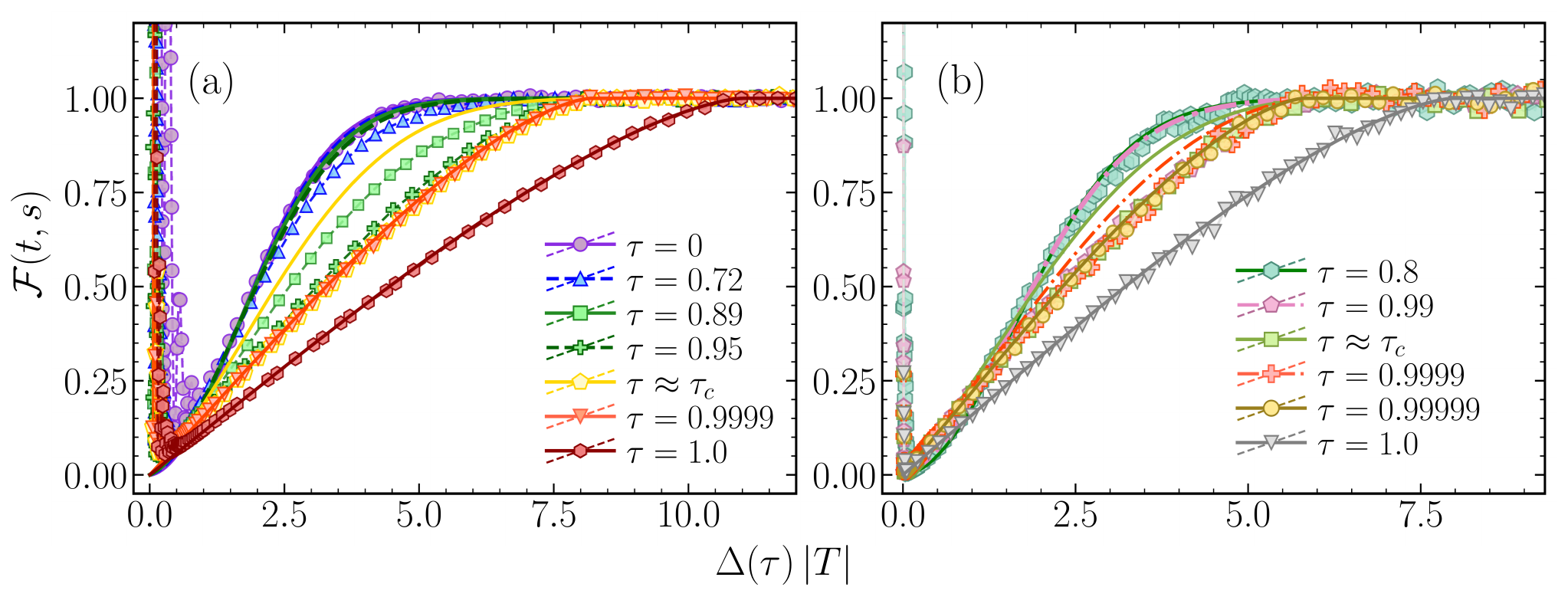}
\caption{Comparison between the DSFF of the cPLBRM model (symbols) and the asymptotic eGinUE analytical results, $1 - \mathcal{F}_{\mathrm{conn}}$ [Eq.~\eqref{FconnAsy}] (solid lines). (a) Datasets correspond to a Hilbert‐space dimension of $128$ with $\varphi=\pi/4$, using $10^{5}$ cPLBRM realizations. The visible mismatch at intermediate $\tau$ values highlights the nonlinear $\tau$–dependence and finite-size effects in cPLBRM at small dimensions, in contrast to the smoother, linear response of the eGinUE ensemble. (b) These datasets correspond to a system dimension of $4096$, with $\varphi=0$, using $4000$ realizations of the cPLBRM ensemble. Here we focus on $\tau$ values in the near-Hermitian regime to highlight the point at which the cPLBRM DSFF becomes insensitive to further changes in $\tau$ and the corresponding curves begin to overlap. In contrast, the eGinUE DSFF continues to evolve with $\tau$ until its spectrum attains a level of non-Hermiticity comparable to that of cPLBRM, after which the two DSFF curves coincide. This behaviour is also evident in panel (a) for smaller system sizes.}
\label{figAppen4}
\end{figure}

Interestingly, we find that with increasing $\tau$, the variance of the imaginary part of the diagonal elements decreases in a nonlinear fashion relative to the overall variance scale, as confirmed by both analytical and numerical analysis. In the regime of small $\tau$, there is a rapid decay of non-Hermitian effects. However, the decay is relatively slow as one approaches $\tau=1$. The behaviour of relative variance is system-size dependent.  In small systems, diagonal imaginary terms contribute noticeably to the total variance and drive the DSFF rapidly toward the Hermitian regime as $\tau$ increases. However, in larger systems, their influence is diluted by the growing dominance of the off-diagonal terms. In large systems, $\tau$ modifies the spectrum only weakly, yielding a smoother evolution of the DSFF. This is in contrast to eGinUE and cSYK, where the variance of all entries scales linearly with $\tau$, making the evolution of DSFF uniform and independent of system size. This linear behavior, when compared with the nonlinear $\tau$–dependence in cPLBRM, produces a visible mismatch between the eGinUE analytics and the cPLBRM simulations at small dimensions within the crossover regime, except at the two extreme values of $\tau$. Figure~\ref{figAppen4}(a) highlights this contrast for small dimensions, while at larger dimensions these finite-size effects diminish and the two ensembles begin to coincide, consistent with the convergence seen in Fig.~\ref{figplbrm} of the main text.

As $\tau$ approaches the Hermitian limit, the imaginary contribution of the diagonal elements in cPLBRM becomes negligible, causing the spectral correlations to saturate and the DSFF of cPLBRM to show minimal sensitivity to $\tau$ once $\tau \gtrsim \tau_c$. Beyond this point, further increase in $\tau$ produce almost no visible change in the DSFF, resulting in curves that largely overlap with each other. In this near-Hermitian regime, however, the eGinUE DSFF continues to evolve until the effective degree of non-Hermiticity becomes comparable to that of cPLBRM. Once this condition is met, the DSFF of cPLBRM and that of eGinUE begin to coincide, showing excellent agreement. This behaviour is illustrated in Fig.~\ref{figAppen4}(b) for a Hilbert-space dimension of 4096, with similar trends observed across other system sizes, such as the dimension 128 case shown in Fig.~\ref{figAppen4}(a).

Based on the observations above, we analysed the crossover behaviour of DSFF in the large-dimensional cPLBRM model and compare it with our analytical findings, as shown in Fig.~\ref{figplbrm} in the main text. Larger matrices suppress finite-size effects and improve agreement with theory but also obscure the crossover regime, as noted at the beginning of this section.

\end{document}